\documentclass[12pt]{article}

\usepackage{amssymb,amsmath,amsfonts,amsthm,enumitem,eurosym,geometry,graphicx,caption,color,setspace,comment,footmisc,pdflscape,array,hyperref,bm,multirow,arydshln,subcaption,placeins,adjustbox,xcolor,float,comment, booktabs}
\usepackage[flushleft]{threeparttable}

\newtheorem{theorem}{Theorem}

\newtheorem{remark}{Remark}

\usepackage[natbibapa]{apacite} 

\hypersetup{%
   colorlinks=true,       
   citecolor=blue,          
   urlcolor  = blue,
   linkcolor=blue,          
}

\title{Estimating Discrete Choice Demand Models with Sparse Market--Product Shocks \footnote{\footnotesize{We would like to thank the seminar participants of Bank of Canada, as well as the 58th Annual Meetings of the Canadian Economics Association, 2024 Econometric Society North American Summer Meeting, IAAE 2024 Annual Conference, and 2025 NSF/NBER Seminar on Bayesian Inference in Econometrics and Statistics for helpful comments. The views expressed in this paper are the authors' and do not reflect those of the Bank of Canada. We gratefully acknowledge financial support from Social Sciences and Humanities Research Council Insight Development Grant.}}
}

\author{Zhentong Lu \and Kenichi Shimizu\footnote{Lu: Financial Stability Department, Bank of Canada, Ottawa, K1A 0G9, Canada. Email: \href{mailto:zlu@bankofcanada.ca}{zlu@bankofcanada.ca}. Shimizu: Department of Economics, University of Alberta, Edmonton, T6G 2H4, Canada. Email: \href{mailto:kenichi.shimizu@ualberta.ca}{kenichi.shimizu@ualberta.ca}.}}

\date{\today}

\begin{document}
\maketitle

\begin{abstract}
\noindent When credible instruments are scarce or estimates hinge on instrument choice, an alternative to the leading Berry--Levinsohn--Pakes (BLP) approach to demand estimation with aggregate data can be useful. We propose a Bayesian random-coefficients discrete-choice demand estimator that jointly recovers preference parameters and market--product demand shocks under a sparsity restriction, without requiring instrumental variables. Shrinkage priors select a sparse set of active shocks, and the posterior supports inference on elasticities, forecasts, and other counterfactuals. Applications to supermarket scanner and automobile data find substantial sparsity and interpretable latent demand variation. We establish identification under sparsity and corroborate the approach with Monte Carlo experiments.

\noindent\textbf{Keywords:} Demand Estimation, Sparsity, Bayesian Inference, Shrinkage Prior

\noindent\textbf{JEL Codes:} C1, C3, L00, D1

\end{abstract}

\section{Introduction}
Demand estimates are central inputs into managerial decisions such as pricing, promotion
planning, assortment design, and product positioning. In differentiated-product markets,
managers often observe rich aggregate data on prices, quantities, product characteristics,
stores, and time periods, but not the latent demand shocks that affect local sales.

A prominent feature of the BLP framework is the market--product demand shock
\(\xi_{jt}\), often interpreted as an unobserved product characteristic observed by firms but
not by the econometrician. This interpretation provides a natural way to model price
endogeneity, since prices may depend on demand shocks observed by firms. It also creates a
difficult identification problem. The demand system alone is underidentified: without
additional restrictions, many combinations of preference parameters and shocks can
rationalize the same observed shares. Standard BLP addresses this problem by inverting the
demand system to recover implied shocks for a candidate parameter value and using excluded
instruments to form GMM moment conditions. The credibility of the resulting estimates
therefore depends on finding instruments that are both relevant and valid.

We illustrate this sensitivity in a familiar setting using the updated U.S. automobile data
of \cite*{grieco2024}, for the industry originally studied in \cite{berry1995automobile}. We
estimate a stylized random-coefficients logit model with a random coefficient on price using
standard two-step BLP GMM and several common instrument sets.\footnote{We use their
data for illustration only and do not attempt to replicate their full demand specification.}
Table~\ref{tab:intro_iv_sensitivity} shows that the estimated mean and standard deviation
of the price coefficient vary substantially across instrument sets, leading to large differences
in implied own-price elasticities. In some specifications, including those based on cost
shifters or BLP-style instruments, the estimated heterogeneity in price sensitivity is
essentially zero, so the random-coefficients model effectively collapses to a simple logit
model. These patterns are consistent with the weak-instrument concerns emphasized by
\cite*{reynaert2014improving}.

\FloatBarrier
\begin{table}[t]
\centering
\caption{Automobile Demand Estimates Across IV Sets}
\label{tab:intro_iv_sensitivity}
\footnotesize
\resizebox{0.9\linewidth}{!}{%
\begin{threeparttable}
\begin{tabular}{l*{6}{>{\centering\arraybackslash}c}}
\hline
 & \shortstack{Cost\\(Ex.\ Rate)} & \shortstack{BLP\\IVs} & \shortstack{Lagged\\Price} & \shortstack{Lagged\\Chars.} & \shortstack{Lag.\ Price\\(Panel)} & \shortstack{Lag.\ Chars.\\(Panel)} \\
\hline
\multicolumn{7}{c}{\textit{Random coefficient on price}} \\
\multirow{2}{*}{\quad Mean} & $-1.21$ & $-1.22$ & $-0.51$ & $-0.69$ & $-0.86$ & $-0.21$ \\
 & $(0.22)$ & $(0.14)$ & $(0.02)$ & $(0.19)$ & $(0.31)$ & $(0.45)$ \\
\multirow{2}{*}{\quad S.D.} & $0.00$ & $0.00$ & $0.21$ & $0.00$ & $0.39$ & $0.00$ \\
 & $(>10^{3})$ & $(>10^{3})$ & $(0.04)$ & $(>10^{3})$ & $(0.32)$ & $(>10^{3})$ \\
\hline
\multicolumn{7}{c}{\textit{Own-price elasticity (across products)}} \\
\quad Mean & $-4.35$ & $-4.40$ & $-1.29$ & $-2.48$ & $-1.65$ & $-0.77$ \\
\quad S.D. & $2.07$ & $2.09$ & $0.31$ & $1.18$ & $0.42$ & $0.37$ \\
\hline
\end{tabular}
\begin{tablenotes}
\scriptsize
\item \textit{Notes:} Automobile data from \cite*{grieco2024}. Two-step BLP GMM with one random coefficient on price; product characteristics and year fixed effects are included. Standard errors (S.E.) in parentheses. When the estimated S.D.\ is at the boundary ($\hat\sigma=0$), the conventional S.E.\ is extremely large (and may not be asymptotically valid); we report these as $(>10^{3})$ rather than the exact numerical value. \textit{Cost (Ex.\ Rate):} lagged real exchange-rate cost shifter and its square. \textit{BLP IVs:} same-firm and rival sums of characteristics. \textit{Lagged Price} and \textit{Lagged Chars.:} own lagged price (level and square) and own lagged product characteristics. Panel columns use the same excluded instruments, with moments based on year-to-year changes in inverted demand shocks within make--model. More details are in Section~\ref{sec:applications}.
\end{tablenotes}
\end{threeparttable}%
}
\end{table}

The instability documented in Table~\ref{tab:intro_iv_sensitivity} motivates an alternative identifying restriction that does not depend on excluded instruments. We work directly with
the aggregate multinomial likelihood and treat the market--product shocks \(\xi_{jt}\) as unknown parameters, estimated jointly with the preference parameters \(\vartheta\), rather than as residuals used to form IV moments. Identification comes from sparsity: we write \(\xi_{jt}=\bar\xi_t+\eta_{jt}\), where \(\bar\xi_t\) is common to all products in market \(t\) and most product-level deviations \(\eta_{jt}\) are zero---as in store-week demand, where only a few products receive local promotional shocks. Building on the spark-based uniqueness logic of \cite{DonohoElad2003}, we show that preference parameters and the shock vector are identified from the aggregate choice likelihood without IV moments, provided sparsity is substantial: fewer than half of the products in a market can carry idiosyncratic shocks. Because \(\xi_{jt}\) is recovered as a parameter, classical price endogeneity is not assumed away: the association between prices and the estimated shocks can be examined ex post.

We implement this idea with a Bayesian shrinkage estimator. The aggregate multinomial
likelihood is combined with global--local shrinkage priors on the product-level deviations
\(\eta_{jt}\). The prior plays the role of variable selection: most deviations are pulled toward
zero, while a small number of active product--market shocks are allowed to remain large
when supported by the data. Preference parameters, market-level components, and active
deviations are updated jointly in a single posterior simulation. The procedure avoids demand
inversion, does not require excluded instruments, and is straightforward to implement with
standard MCMC tools.

This turns the estimation problem into a high-dimensional variable-selection problem. While shrinkage
priors are widely used to regularize large parameter spaces, including in Bayesian
macroeconomic forecasting \citep{giannone2015prior,GiannoneLenzaPrimiceri2021},
their role in nonlinear aggregate discrete-choice likelihoods is less developed; here the
prior both regularizes estimation and implements the sparsity restriction used for
identification.

Our approach also connects to Bayesian likelihood-based estimation of aggregate demand (\citealp{YangChenAllenby2003bayesian} and
\citealp{MusalemBradlowRaju2009BayesBLP}). \cite{berry2003comment} noted that treating product-level demand shocks as parameters requires explaining how they are separately identified from preference parameters. Sparsity provides such an identifying restriction.

Treating \(\xi_{jt}\) as a parameter also simplifies inference for decision-relevant counterfactuals.
In standard BLP applications, price elasticities are typically evaluated at the point estimates
\((\hat\vartheta,\hat\xi)\), with uncertainty obtained, if at all, by propagating sampling variation in
\(\hat\vartheta\) through GMM standard errors or bootstrap methods. Uncertainty in the recovered
demand shocks \(\xi\) is rarely incorporated. In our framework, price elasticities, predicted shares,
and other counterfactual quantities are functions of the joint parameter vector
\((\vartheta,\xi)\), so their uncertainty is obtained directly from posterior draws.

The same framework naturally extends to prediction. Because standard BLP recovers demand
shocks only for observed markets, forecasting demand in a new store, week, or model year
requires an additional rule for the unobserved shock. In contrast, our posterior predictive
distribution provides a coherent forecasting rule while shrinkage regularizes the latent demand
component, improving the tradeoff between in-sample fit and out-of-sample prediction.

Because $\eta_{jt}$ is estimated rather than treated as a residual, its posterior provides a diagnostic measure of unusual product–market outcomes. Shrinkage is essential for this interpretation because it prevents latent shocks from simply absorbing residual variation. In scanner data, large positive deviations may indicate unusually strong promotional execution, shelf placement, or local merchandising effort, while large negative deviations may reflect unusually weak support (e.g., poor shelf location or limited visibility relative to peers). In durable-goods markets, they may capture model-year appeal, advertising, or product
positioning not fully explained by observed characteristics. 

We illustrate the method using yogurt scanner data and the U.S. automobile market. The yogurt application shows that the estimated sparse deviations recover omitted promotional activity, while the automobile application revisits the IV-sensitive setting from
Table~\ref{tab:intro_iv_sensitivity} and demonstrates economically plausible demand estimates and competitive predictive performance.


The rest of the paper is organized as follows. Section~\ref{sec:model} presents the demand model, the
sparsity restriction on market--product shocks, and the Bayesian shrinkage estimator.
Section~\ref{sec:applications} studies the yogurt and automobile applications. Section~\ref{sec:identification} establishes
identification under sparsity, compares this source of identification with the conventional
BLP--IV approach, and presents Monte Carlo evidence. Section~\ref{sec:conclusion} concludes.

\subsection{Related Literature}
This paper first relates to the large literature on identification and estimation in BLP-type
demand models. As emphasized by \cite*{BerryHaile2014ECMA}, standard BLP estimation
relies on excluded instruments for identification. Many instruments have been proposed,
including 
cost shifters (\citealp*{berry1999voluntary, goldberg2001evolution}), 
BLP IVs (\citealp*{berry1995automobile}), 
Hausman IVs (\citealp*{hausman1994valuation, nevo2001measuring}), 
optimal IVs (\citealp*{berry1999voluntary, reynaert2014improving}), 
differential IVs (\citealp*{gandhi2019measuring}), and 
time-series or panel data-based IVs (\citealp*{sweeting2013dynamic, jin2021flagship}). 
Our paper is motivated by settings in which estimates are sensitive to these choices, potentially because of weak instruments \citep*{armstrong2016large}. Instead of using excluded IV moments, we impose sparsity on market--product demand shocks as an alternative identifying restriction. 
This identification strategy is also related to the broader literature on identification of random-coefficients discrete-choice models (\citealp*{fox2012random}, \cite*{fox2016nonparametric}, \citealp*{lu2023semi}, \citealp*{dunker2023nonparametric}). 
Our setting differs in that market--product demand shocks are treated as unknown parameters, with sparsity providing the identifying restriction. 
The objective is  to recover both the random-coefficient distribution and the sparse demand shocks from aggregate shares.

Several papers impose structure on demand shocks or use additional data to reduce reliance
on conventional IVs. \cite*{MoonShumWeidner2018blp} model demand shocks using interactive
fixed effects, \cite*{GillenMonteroMoonShum2019blp} use LASSO-type methods to select rich
controls in BLP, and \cite*{ByrneImaiJainSarafidis2022ivfree} use cost data for identification
without requiring cost variables to be valid demand instruments. Our approach instead identifies the model through sparsity in demand shocks, requires no cost-side data, and estimates all parameters jointly through a Bayesian shrinkage posterior.

The paper also contributes to Bayesian approaches to  aggregate demand estimation.
\cite*{Jiang2009bayesian}  develop a Bayesian BLP procedure in which the likelihood is
constructed through demand inversion. In contrast, we treat market--product demand shocks
as parameters in the aggregate likelihood and estimate them jointly with preference
parameters.  
This avoids demand inversion and identifies the model through sparsity, with shrinkage priors providing a practical implementation.  
More broadly, our use of shrinkage to select sparse market--product shocks connects to high-dimensional demand estimation, including random projections \citep*{ChiongShum2019MS}, random partitions \citep*{SmithAllenby2019JASA}, scalable high-dimensional choice models \citep*{LoaizaNibbering2022ScalableProbit_JBES,Jiang2024high_MS}, demand for bundles \citep*{WangIaria2024_JEEA}, complementarity with many products \citep*{Ershov2024RAND}, and scalable consideration-set models \citep*{ChibShimizu2026scalable}.

\section{Model and Estimation} \label{sec:model}
\subsection{Demand Model}
We consider a stylized random coefficient logit demand model for aggregate data, in the spirit of \cite*{berry1995automobile}. There are \(T\) markets, indexed by \(t = 1, \dots, T\), each consisting of \(J_t + 1\) products, indexed by \(j = 0, 1, \dots, J_t\), and \(N_t\) consumers, indexed by \(i = 1, \dots, N_t\). The products indexed by \(j > 0\) are ``inside goods,'' and product \(0\) is the ``outside option.'' 

Each consumer \(i\)'s utility from product \(j\) in market \(t\) is given by
\begin{equation}
u_{ijt} = X_{jt}^{\top} \beta_i + \xi_{jt} + \varepsilon_{ijt}, \label{eq:utility}
\end{equation}
where \(X_{jt} \in \mathbb{R}^{d_X}\) is a vector of observed market--product characteristics, \(\beta_i\) represents consumer-specific taste parameters (i.e., random coefficients), which are i.i.d. across consumers with distribution \(f(\cdot \mid \vartheta)\), where \(\vartheta \in \Theta \subset \mathbb{R}^{d_\vartheta}\) is a finite-dimensional parameter vector, \(\xi_{jt}\) is the market--product level demand shock (a.k.a. unobserved characteristic), and \(\varepsilon_{ijt}\) is an i.i.d. idiosyncratic preference shock across \(i\), \(j\), and \(t\), following the standard Gumbel distribution. To normalize the level of the random utility, the product characteristics and demand shock of the outside option, \(X_{0t}, \xi_{0t}\), are set to zero. 
As in standard BLP models, some variables in \(X_{jt}\) such as price, may be endogenous because firms observe \(\xi_{jt}\) when setting them. We discuss this issue further in Section \ref{sec:identification}.

The implied  share of product $j$ in market $t$ is
\begin{equation} \label{eq:mk_sh_model_pred}
\sigma_{jt}\left(\xi_t, \vartheta\right) = \int \frac{\exp\left(X_{jt}^{\top} \beta + \xi_{jt}\right)}{1 + \sum_{k=1}^{J_t} \exp\left(X_{kt}^{\top} \beta + \xi_{kt}\right)} f(\beta \mid \vartheta) \, d\beta,
\end{equation}
where we denote the \(J_t\)-dimensional vector by 
\(\xi_t = (\xi_{1t}, \dots, \xi_{J_t t})^{\top}\).

Let $q_{jt}$ denote the observed aggregate quantity sold for product $j$ in market $t$. The corresponding market shares are $s_{jt}=q_{jt}/N_t$, and the aggregate multinomial likelihood is 
\begin{equation} \label{eq:likelihood_agg} L\left(\vartheta,\xi_{1},...,\xi_{T}\right)
=\prod_{t=1}^{T}\prod_{j=0}^{J_{t}}\left[\sigma_{jt}\left(\xi_{t},\vartheta\right)\right]^{q_{jt}}.
\end{equation}
When individual-level data are available, our approach can be modified easily to incorporate this information.\footnote{Such an extension is available upon request.}
Because there are \(\sum_{t=1}^{T}J_{t}+d_{\vartheta}\) unknown parameters but only \(\sum_{t=1}^{T}J_{t}\) independent demand equations, the system 
\begin{equation} \label{eq:demand_system} 
s_{jt}=\sigma_{jt}\left(\xi_{t},\vartheta\right), \:\forall j,t,
\end{equation}
is underidentified without additional restrictions.

\subsection{Sparse Demand Shocks}\label{subsec:sparse_xi}

To address the underidentification problem in (\ref{eq:demand_system}), the standard BLP approach \cite{berry1995automobile} recovers  \(\xi_{jt}\) by inverting the demand system 
and estimates \(\vartheta\) using GMM with excluded instruments. We instead impose a sparse structure on \(\xi_{jt}\), treating the shocks as parameters in the aggregate likelihood and estimating them jointly with \(\vartheta\).

We consider the within-market decomposition
\begin{equation}
    \xi_{jt}=\bar{\xi}_t+\eta_{jt}, \label{eq:xi}
\end{equation}
where \(\bar{\xi}_t\) is a market-level shock common to all inside products in
market \(t\) and \(\eta_{jt}\) is a product-level deviation. Sparsity applies to \(\{\eta_{jt}\}\): for most products, \(\eta_{jt}=0\) and \(\xi_{jt}=\bar{\xi}_t\); only a minority of products carry a nonzero deviation from the common market shock. The effective number of distinct shocks per market is therefore \(1\) plus the number of active deviations, which is much smaller than \(J_t\) when sparsity is strong. More flexible sparse patterns---such as ties of \(\xi_{jt}\) across products or markets beyond (\ref{eq:xi})---could be accommodated with the same shrinkage logic, but we leave them for future work.\footnote{%
For example, one could allow arbitrary partitions of product--market pairs with
common shock values within each cell, including ties across markets. Related BLP approaches that impose low-dimensional structure on demand shocks
include \cite{MoonShumWeidner2018blp} and \cite{GillenMonteroMoonShum2019blp}.}

Selecting which \(\eta_{jt}\) are active is a high-dimensional variable-selection problem. As in high-dimensional regression, penalized estimators such as the LASSO and Bayesian global–local or spike-and-slab priors shrink most coefficients toward zero while allowing a small subset to remain active.\footnote{See, e.g.,
\cite{Tibshirani1996lasso}, \cite{KyungGillGhoshCasella2010}, and
\cite{polson2010shrink} for penalized and Bayesian shrinkage methods;
\cite{KorobilisShimizu2022} surveys applications in econometrics.} 

\subsection{Bayesian Shrinkage Approach to Estimation}\label{sec:bayes_estimation}
We implement the sparse specification (\ref{eq:xi})  using Bayesian shrinkage priors. 
The aggregate multinomial likelihood is combined with shrinkage priors on $\eta$, which simultaneously select active deviations and estimate their magnitudes while providing posterior uncertainty quantification.

\subsubsection{Likelihood}\label{subsec:likelihood}
We assume independent normal random coefficients,\footnote{\cite*{berry1995automobile} also impose independence across random coefficients.}
\[
\beta \sim N_{d_X}(\bar{\beta},\Sigma),  \quad 
\Sigma=\text{diag}(\sigma^2_{1},\ldots,\sigma^2_{d_X}).
\]
Following \cite*{Jiang2009bayesian}, we reparametrize \(\Sigma\) through the log standard deviations: \(r_k=\log(\sigma_k)\). 
Collect the parameters entering the likelihood as \(\theta=(\bar{\beta}, r, \bar{\xi}, \eta)\), with \(\xi_{jt}=\bar{\xi}_t+\eta_{jt}\) from (\ref{eq:xi}). 
The log-likelihood is, up to the multinomial coefficient which does not depend on \(\theta\),
\begin{equation}\label{eq:likelihood_estimation}
\ell(\theta)
=
\sum_{t=1}^T 
\left\{ 
q_{0t} \log \sigma_{0t}(\theta)
+
\sum_{j=1}^{J_t} q_{jt} \log \sigma_{jt}(\theta)
\right\},
\end{equation}
where the predicted shares are  approximated using 
\(R_0\) i.i.d.\ \(N(0,1)\) draws \(v_r\): 
\[
\sigma_{jt}(\theta)
\approx
\frac{1}{R_0}\sum_{r=1}^{R_0}
\frac{\exp\left(\delta_{jt} + \mu_{jt}(v_r) \right)}
{1+\sum_{k=1}^{J_{t}}\exp\left(\delta_{kt} + \mu_{kt}(v_r) \right)},
\qquad j=1,\ldots,J_t,
\]
where 
\(\delta_{jt}=X_{jt}'\bar{\beta}+\xi_{jt}\), 
\(\mu_{jt}(v_i)=X_{jt}'R v_i\)
with $R=\text{diag}(\sigma_1,\ldots,\sigma_{d_X})$, 
and 
$
\sigma_{0t}(\theta)=1-\sum_{j=1}^{J_t}\sigma_{jt}(\theta). 
$


\subsubsection{Prior}\label{sec:prior}
We employ the horseshoe prior of \cite{carvalho2010horseshoe} on the market--product deviations:
\begin{equation}
    \eta_{jt}\sim N\left(0, \tau^2_0 \lambda^2_{jt}\right), \ j=1,\ldots,J_t, \ t=1,\ldots,T, 
\end{equation}
with half-Cauchy priors on both 
local scales  $ \lambda_{jt}$ and 
the global scale $\tau_0$. 
The horseshoe aggressively shrinks small signals toward zero while leaving large signals relatively unshrunk because of its heavy tails.\footnote{An attractive feature of the horseshoe prior is that it  requires less tuning than Lasso-type shrinkage methods, since the global and local shrinkage levels are learned from the data rather than selected through procedures such as cross-validation.
Numerous theoretical results have been established for this prior. 
See \cite{datta2013asymptotic} and \cite{van2014horseshoe} for a more detailed discussion.} 
For the remaining parameters, we specify independent priors:
\begin{align*}
    \bar{\beta} &\sim N_{d_X}(\underline{\mu}_\beta,\underline{V}_\beta),\\
    \bar{\xi}_t &\overset{ind}{\sim} N(\underline{\mu}_{\xi_t},\underline{V}_{\xi_t}),  \quad  t=1,\ldots,T,\\
    r_k&\overset{ind}{\sim} N(0,\underline{V}_{r,k}) ,    \quad   k=1,\ldots,d_X.
\end{align*}
We let 
$(\underline{\mu}_\beta,\underline{V}_\beta)=(0, 10 \cdot I_{d_X})$ and 
$(\underline{\mu}_{\xi_t},\underline{V}_{\xi_t})=(0, 10)$ for all markets $t$ to give sufficiently  uninformative priors on the fixed slope parameter $\bar{\beta}$ and the market-specific shocks $\bar{\xi}_t$.  
For the log standard deviations, we let 
$\underline{V}_{r,k}=0.5$ for all $k$ to give weakly informative priors on $\Sigma$.

\subsubsection{Posterior inference}

%
Combining the likelihood \eqref{eq:likelihood_estimation} with the priors yields the posterior
\begin{equation}
     p\left( \bar{\beta},  r,  \bar{\xi}, \eta,  \lambda, \tau_0 \vert  q \right)
    \propto 
    p\left(  q \vert  \bar{\beta},  r,  \bar{\xi}, \eta \right)
    \cdot
    p\left(  \bar{\beta},  r,  \bar{\xi}, \eta,  \lambda, \tau_0 \right).\label{eq:JointPosterior}   
\end{equation}
Posterior simulation is performed in \texttt{stan} using Hamiltonian Monte Carlo. 
Posterior draws provide inference for price elasticities and other counterfactual quantities.

\begin{remark}
Unlike existing Bayesian BLP approaches, our estimator works directly with the aggregate likelihood without repeated demand inversion. This simplifies computation, naturally accommodates zero market shares \citep*{gandhi2023estimating}, and yields posterior uncertainty for elasticities and other counterfactual quantities.




\end{remark}

\section{Empirical Evidence on Sparse Demand Shocks}\label{sec:applications}
We examine whether the sparse demand structure is empirically relevant in two settings:
store-level yogurt demand and the U.S. automobile market. The applications differ in
industry context, market definition, and product frequency. In the yogurt data, observed
promotion variables are omitted from utility and used to validate the estimated product-level
deviations. The automobile application revisits the IV-sensitive setting discussed in Section 1.

In both applications, we compare three specifications: (1) no product-level deviations,
$\eta_{jt}=0$ (no-$\eta)$; (2) the shrinkage-prior specification; and (3) a diffuse-prior benchmark,
$\eta_{jt}\sim N(0,10^2)$. 
The no-$\eta$ specification restricts all product deviations to zero, so unobserved demand variation is captured only by observed covariates and random coefficients. The diffuse specification allows product deviations to vary freely. The shrinkage specification lies between these two extremes by allowing non-zero deviations only when supported by the data.

As the global--local shrinkage prior introduced in Section \ref{sec:prior} is continuous,  the posterior estimates of $\eta_{jt}$ are never exactly zero.  In order to summarize sparsity ex-post, 
we classify a deviation as economically negligible when the absolute
value  of its posterior mean falls below a calibrated threshold i.e.\ $\vert \hat{\eta}_{jt} \vert <\epsilon$. The threshold is
chosen so that a utility perturbation of size $\epsilon$ changes predicted market share by
0.1 percentage points on average; see Appendix~\ref{sec:calibration_threshold}. 
We use $R_0=50$ simulation draws and 6,000 MCMC draws,
discarding the first 1,000 as burn-in.


\subsection{Yogurt Demand and Sparse Promotions}

\subsubsection{Data and Specification}

We use IRI scanner data for yogurt products in 95 stores in the New York metropolitan area (as defined by IRI) during the week of June 25--July 1, 2012.\footnote{See \cite*{bronnenberg2008iri} for a description of the IRI marketing data set.} The data contain UPC-level information on prices, quantities, product characteristics, and promotional activities for each store.

We define a ``product'' by aggregating UPCs with the same brand, size category, and characteristics (flavor, fat content, Greek, and organic status). Prices are quantity-weighted averages across UPCs within a product; quantities are summed.

A ``market'' is a store in the sample week, and the choice set is the set of products available in that store. Market shares equal quantities sold divided by the local population (within a 2-mile radius of the focal store) in the IRI data, yielding 5,927 product--store observations. Table~\ref{tab:sum_stats_yg_sample} reports summary statistics for several products. 
Mean utility includes price, brand fixed effects, size indicators, and the product
characteristics reported in Table~\ref{tab:sum_stats_yg_sample}, with random coefficients on price and organic dummy.

Conditional on these controls and the store-level component $\bar{\xi}_t$, $\eta_{jt}$
is the product-specific deviation within a store. We interpret positive deviations as
unusually strong promotional or merchandising support, such as display activity or
preferential shelf placement, and negative deviations as unusually weak support.
Because only a limited subset of products is likely to experience such conditions in a
given store and week, the deviations are plausibly sparse. We omit \textit{Display} and
\textit{Feature} from utility and reserve them for ex-post validation of $\hat{\eta}_{jt}$.

To assess  predictive performance of the estimators, we randomly partition the stores into five folds.
For each fold, we estimate the model using four folds and evaluate predictive
performance on the remaining held-out fold. We average the results across the
five folds. Because product-level deviations are unobserved in held-out stores,
we draw them from the corresponding posterior predictive distribution.

\begin{table}[h!]
\caption{Summary statistics for several randomly chosen products}
\label{tab:sum_stats_yg_sample}
\centering
\resizebox{\linewidth}{!}{%
\begin{threeparttable}
\begin{tabular}{ccccccccccccc}
\hline 
Product  & No. of  & Market & \multirow{2}{*}{Price} & \multicolumn{7}{c}{Product Characteristics} & \multicolumn{2}{c}{Marketing Mix}\tabularnewline
\cline{5-13} \cline{6-13} \cline{7-13} \cline{8-13} \cline{9-13} \cline{10-13} \cline{11-13} \cline{12-13} \cline{13-13} 
No. & Markets & Share (\%) &  & Brand & Size (pt.) & Flavor & Low Fat & No Fat & Greek & Organic & Display & Feature\tabularnewline
\hline 
1 & 10 & 0.015 & 2.459 & ALPINA & 0.4 & 0 & 1 & 0 & 0 & 0 & 0 & 0\tabularnewline
 &  & (0.009) & (0.627) &  &  &  &  &  &  &  &  & \tabularnewline
2 & 63 & 2.575 & 1.348 & AXELROD & 0.375 & 1 & 1 & 0 & 0 & 0 & 0.090 & 0.787\tabularnewline
 &  & (2.024) & (0.393) &  &  &  &  &  &  &  & (0.222) & (0.407)\tabularnewline
3 & 20 & 0.026 & 1.482 & AXELROD & 2 & 0 & 0 & 1 & 0 & 0 & 0.044 & 0\tabularnewline
 &  & (0.027) & (0.099) &  &  &  &  &  &  &  & (0.211) & \tabularnewline
4 & 90 & 1.370 & 3.051 & CHOBANI & 0.375 & 0 & 1 & 0 & 1 & 0 & 0.219 & 0.555\tabularnewline
 &  & (1.362) & (0.533) &  &  &  &  &  &  &  & (0.416) & (0.500)\tabularnewline
5 & 52 & 0.015 & 3.643 & CHOBANI & 1 & 1 & 1 & 0 & 1 & 0 & 0 & 0.028\tabularnewline
 &  & (0.019) & (0.411) &  &  &  &  &  &  &  &  & (0.166)\tabularnewline
6 & 58 & 0.053 & 3.033 & CHOBANI & 2 & 1 & 1 & 0 & 1 & 0 & 0.002 & 0.011\tabularnewline
 &  & (0.055) & (0.306) &  &  &  &  &  &  &  & (0.028) & (0.104)\tabularnewline
7 & 51 & 0.069 & 2.211 & YOPLAIT & 0.375 & 1 & 1 & 0 & 0 & 0 & 0.042 & 0.144\tabularnewline
 &  & (0.071) & (0.318) &  &  &  &  &  &  &  & (0.202) & (0.354)\tabularnewline
8 & 85 & 0.264 & 2.120 & YOPLAIT & 0.375 & 1 & 0 & 1 & 0 & 0 & 0.030 & 0.104\tabularnewline
 &  & (0.263) & (0.263) &  &  &  &  &  &  &  & (0.157) & (0.307)\tabularnewline
\hline 
\end{tabular}

\begin{tablenotes}
\item Note: This table presents summary statistics for a selection of randomly chosen products in the yogurt category. The first number in each cell represents the mean, while the second number in parentheses indicates the standard deviation across markets. If the standard deviation is zero, it is omitted. The ``No. of Markets'' column indicates the markets where each product is available, highlighting substantial variations in consumers' choice sets. The ``Market Share (\%)'' and ``Price'' columns, as well as the marketing mix columns, report the averages across different markets for each product. 
\end{tablenotes}
\end{threeparttable}
}
\end{table}

\subsubsection{Estimation Results} 
Table~\ref{tab:estimation_table_yg} compares the no-$\eta$, shrinkage, and diffuse-prior specifications. The shrinkage estimates generally lie between the no-$\eta$ and diffuse-prior results,
but are much closer to the no-$\eta$ specification. For example, the price coefficient is
$-1.474$ under shrinkage, compared with $-1.120$ under no-$\eta$ and $-4.640$ under
the diffuse prior. The diffuse prior also produces substantially larger posterior standard
deviations, consistent with weak identification when flexible product deviations absorb
variation that would otherwise identify preferences.

The no-$\eta$ and shrinkage specifications imply similar average own-price elasticities
($-2.36$ and $-2.34$), whereas the diffuse prior produces a much larger and more dispersed
distribution, with an average of $-8.94$ and a minimum of $-24.25$. These extreme values
are concentrated among products with very small observed shares (see Supplementary Material Figure~\ref{fig:yg_elasticity_share}), consistent with overfitting under the diffuse prior.\footnote{For reference, existing studies of the U.S.\ yogurt market based on the BLP GMM approach report average own-price elasticities of around $-5$ \citep{villasboas2007vertical, draganska2006consumer}. Our estimates are not directly comparable, as they differ in data and aggregation method.}

Shrinkage substantially improves in-sample fit relative to no-$\eta$:
the RMSE falls from 0.0046 to 0.0016, compared with 0.0002 under the diffuse prior. The corresponding share-fit plots can be found in Supplementary Material Figure~\ref{fig:share_fit_compare_yg}. 

Out of sample, no-$\eta$ has the lowest holdout RMSE (0.0089), followed by shrinkage (0.0161) and the diffuse prior (0.0264).\footnote{When held-out deviations
are set to zero for all specifications, shrinkage nearly matches no-$\eta$; see Table~\ref{tab:estimation_table_yg_appdx}.} Thus, the shrinkage estimator achieves a better balance between in-sample fit and predictive performance than the diffuse specification.

Unlike the no-$\eta$ specification, shrinkage also recovers a sparse set of
product--store deviations. We next examine whether these deviations capture economically meaningful information. Specifically, we check whether they align with
recorded \textit{Display} and \textit{Feature} activity that we intentionally omit from estimation.


\begin{table}[!h]
\centering
\caption{Posterior parameter estimates and elasticities}
\label{tab:estimation_table_yg}
\centering
\resizebox{0.8\linewidth}{!}{
\begin{threeparttable}
\begin{tabular}[t]{lcccccc}
\toprule
\multicolumn{1}{c}{ } & \multicolumn{2}{c}{No $\eta$} & \multicolumn{2}{c}{Shrinkage prior} & \multicolumn{2}{c}{Diffuse prior} \\
\cmidrule(l{3pt}r{3pt}){2-3} \cmidrule(l{3pt}r{3pt}){4-5} \cmidrule(l{3pt}r{3pt}){6-7}
 & Mean & SD & Mean & SD & Mean & SD\\
\midrule
\textbf{Mean Coefficient} & & & & & & \\
Price & -1.120 & 0.030 & -1.474 & 0.020 & -4.640 & 0.232\\
Size 2 & -3.076 & 0.116 & -2.868 & 0.098 & -10.944 & 0.959\\
Size 3 & -5.400 & 0.850 & -4.828 & 0.718 & -10.717 & 1.913\\
Size 4 & -3.062 & 0.051 & -2.915 & 0.062 & -10.574 & 0.524\\
Flavored & 1.642 & 0.035 & 1.175 & 0.034 & 0.961 & 0.446\\
Nonfat & 1.179 & 0.076 & 0.729 & 0.070 & -0.024 & 0.513\\
Lowfat & 0.730 & 0.076 & 0.483 & 0.055 & -0.727 & 0.525\\
Greek & 0.167 & 0.052 & 0.195 & 0.048 & 0.343 & 0.754\\
Organic & -0.543 & 0.130 & -0.918 & 0.155 & -3.106 & 1.041\\
Brand FEs (omitted) \\
\midrule
\textbf{SD of Random Coefficient} & & & & & & \\
$\sigma_{\text{Price}}$ & 0.290 & 0.017 & 0.479 & 0.013 & 0.657 & 0.088\\
$\sigma_{\text{Organic}}$ & 0.517 & 0.211 & 0.804 & 0.174 & 0.937 & 0.498\\

\midrule

\addlinespace[0.3em]
\multicolumn{7}{l}{\textbf{Own Elasticities}}\\
\hspace{1em}Average own elasticity & -2.363 &  & -2.339 &  & -8.935 & \\
\hspace{1em}SD of own elasticity & 0.786 &  & 0.572 &  & 3.659 & \\
\hspace{1em}Minimum own elasticity & -4.797 &  & -4.740 &  & -24.247 & \\
\hspace{1em}Maximum own elasticity & -0.856 &  & -1.032 &  & -2.706 & \\
\addlinespace[0.3em]
\multicolumn{7}{l}{\textbf{Cross Elasticities (omitted)}}\\
\midrule

\addlinespace[0.3em]
\multicolumn{7}{l}{\textbf{In-sample Fit}}\\
\hspace{1em}RMSE & 0.0046 &  & 0.0016 &  & 0.0002 & \\

\addlinespace[0.3em] 
\multicolumn{7}{l}{\textbf{Out-of-sample Prediction}}\\ 
\hspace{1em}RMSE & 0.0089 & & 0.0161 & & 0.0264 & \\ 


\bottomrule
\end{tabular}
\begin{tablenotes}
\item \textit{Note: } The table reports posterior means and posterior standard deviations for model parameters. 
Size 1 is the omitted baseline category.
Brand fixed effects are included but omitted from the table for brevity. The specification includes random coefficients on price and organic. 
Elasticity summaries are computed across products using posterior mean own-price elasticities. The No $\eta$ specification restricts product deviations to zero, so unobserved demand varies only through market-level components. Out-of-sample prediction is evaluated using store-level holdout samples.  
RMSEs are computed across product--store observations in the held-out stores.
Results are based on a random partition of the stores into five folds.
\end{tablenotes}
\end{threeparttable}}
\end{table}

\subsubsection{Interpreting Sparse Demand Shocks}
\begin{figure}[ht]
    \centering

    \begin{subfigure}{0.48\textwidth}
        \centering
        \includegraphics[width=\linewidth]{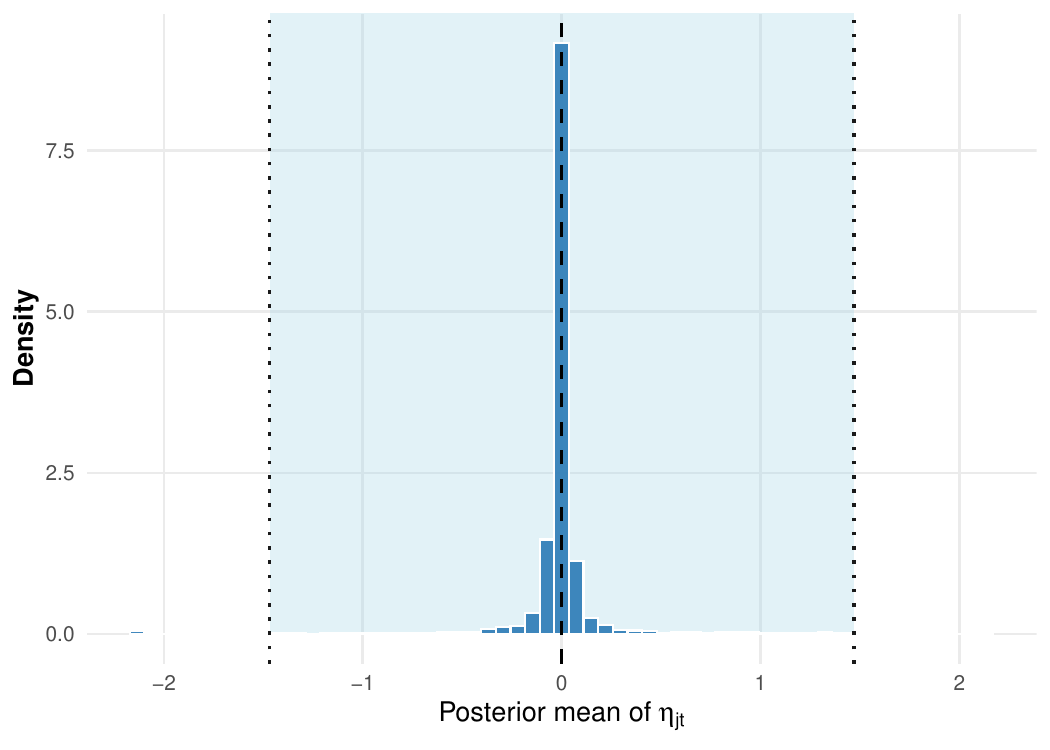}
        \caption{Shrinkage prior}
        \label{fig:yg_eta_hist_shrink}
    \end{subfigure}
    \hfill
    \begin{subfigure}{0.48\textwidth}
        \centering
        \includegraphics[width=\linewidth]{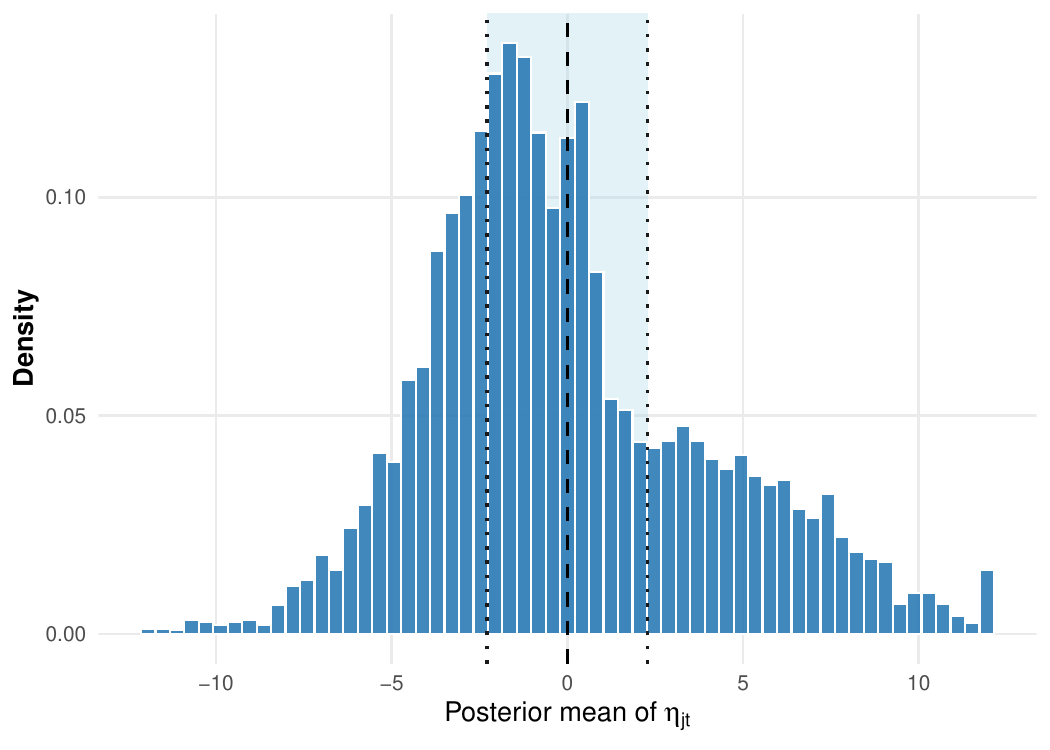}
        \caption{Diffuse prior}
        \label{fig:yg_eta_hist_diffuse}
    \end{subfigure}

\caption{Distribution of posterior means of product-level deviations $\eta_{jt}$. 
The threshold $\epsilon$ is calibrated to correspond to an average 0.1 percentage-point change in predicted market share (See Section \ref{sec:calibration_threshold}  for detail). 
Under the shrinkage prior, with the corresponding diffuse-prior values in
parentheses, the calibrated threshold is $\epsilon=1.469$ (2.282), and
98.7\% (44.7\%) of the posterior mean deviations lie within $\pm\epsilon$.
The corresponding counts are 5,849 of 5,927 (2,648 of 5,927). }
\label{fig:application_yg_eta_hist}
\end{figure}

The posterior mean deviations are strongly concentrated near zero under shrinkage.
Figure~\ref{fig:application_yg_eta_hist} shows that 98.7\% lie within the calibrated threshold, compared with
44.7\% under the diffuse prior.

We use the recorded \textit{Display} and \textit{Feature} indicators for
ex-post validation rather than equating them directly with $\eta_{jt}$.
If the estimated deviations capture unusual promotional or merchandising support,
$\hat{\eta}_{jt}$ should be positively associated with these indicators.

\begin{table}[h!]
\centering
\begin{threeparttable}
\caption{Promotion lift in latent demand deviations}
\label{tab:promotion_lift_eta_yg}
\begin{tabular}{lccccccc}
\toprule
 &  & \multicolumn{3}{c}{Shrinkage prior} & \multicolumn{3}{c}{Diffuse prior} \\
\cmidrule(l{3pt}r{3pt}){3-5} \cmidrule(l{3pt}r{3pt}){6-8}
Promotion & Active share & Inactive & Active & Lift & Inactive & Active & Lift \\
\midrule
Display & 3.8\% & -0.007 & 0.182 & 0.189 & -0.056 & 1.426 & 1.482\\
Feature & 11.3\% & -0.006 & 0.045 & 0.050 & 0.033 & -0.261 & -0.294\\
\bottomrule
\end{tabular}
\begin{tablenotes}
\footnotesize
\item \textit{Notes:} The table reports mean posterior product-level demand deviations $\hat\eta_{jt}$ by promotion status. A promotion is classified as active if its value is strictly positive. The lift is the difference between the mean $\hat\eta_{jt}$ among active product--store observations and the mean among inactive observations.
\end{tablenotes}
\end{threeparttable}
\end{table}

Table~\ref{tab:promotion_lift_eta_yg} supports this interpretation. Under shrinkage, the mean
$\hat{\eta}_{jt}$ is 0.189 higher for products with display activity than
for those without, while the corresponding feature gap is 0.050.
Under the diffuse prior, the display gap is much larger, but the feature gap
becomes negative, which is inconsistent with the proposed promotional
interpretation. See Supplementary Material Table~\ref{tab:eta_marketing_mix_regression_yg} for a similar regression-based check.

Overall, shrinkage recovers a sparse set of product-specific deviations that
align with recorded \textit{Display} and \textit{Feature} activity. The no-$\eta$
specification cannot recover such deviations, while the diffuse prior overfits
market shares, produces more extreme elasticities, and does not reproduce the
same promotional pattern.

\subsection{Automobile market}
We next revisit the U.S. automobile market, using the same data as in the introductory IV-sensitivity exercise in Table~\ref{tab:intro_iv_sensitivity}. This application is useful because the standard BLP estimator is sensitive to the choice of instruments in this setting, with estimates of both the mean and the standard deviation of the price coefficient varying substantially across IV specifications, and with implied price elasticities varying as well. We therefore use the automobile market to ask whether sparsity in product-market demand shocks provides a disciplined alternative source of identification\footnote{
We note that our specification differs from \cite{berry1995automobile} and \cite{grieco2024} in two key ways: (1) it excludes a supply-side model, and (2) it includes market fixed effects to account for market-level heterogeneity. Our goal is not to replicate their results but to illustrate our approach in this classic dataset.}.

\subsubsection{Data and Specification}
Each year is treated as a market. The sample contains 39 markets, spanning 1980--2018, and 9,694 product--market observations.\footnote{The data are from \cite*{grieco2024}; see their paper and replication package for further details on the sample construction and variables.} Our utility specification includes price, vehicle characteristics, category indicators, and make fixed effects. We include a random coefficient on price and decompose unobserved demand
as $\xi_{jt}=\bar{\xi}_t+\eta_{jt}$, as in  (\ref{eq:xi}).

\subsubsection{Estimation Results}

\begin{table}[!h]
\centering
\caption{Posterior parameter estimates and elasticities}
\label{tab:application_estimation_table_at}
\centering
\resizebox{0.8\linewidth}{!}{
\begin{threeparttable}
\begin{tabular}[t]{lcccccc}
\toprule
\multicolumn{1}{c}{ } & \multicolumn{2}{c}{No $\eta$} & \multicolumn{2}{c}{Shrinkage prior} & \multicolumn{2}{c}{Diffuse prior} \\
\cmidrule(l{3pt}r{3pt}){2-3} \cmidrule(l{3pt}r{3pt}){4-5} \cmidrule(l{3pt}r{3pt}){6-7}
 & Mean & SD & Mean & SD & Mean & SD\\
\midrule
\textbf{Mean Coefficient} & & & & & & \\
MSRP (Price) & -0.417 & 0.006 & -0.807 & 0.007 & -3.926 & 0.361\\
Van & -0.355 & 0.012 & -0.224 & 0.013 & 0.113 & 0.134\\
SUV & -0.024 & 0.009 & 0.058 & 0.007 & 0.567 & 0.142\\
Truck & 0.336 & 0.009 & 0.081 & 0.010 & -0.868 & 0.159\\
Log(Footprint) & 1.024 & 0.033 & -0.691 & 0.014 & -0.571 & 0.351\\
Log(HP) & 0.693 & 0.021 & 1.224 & 0.017 & 2.837 & 0.515\\
Log(MPG) & 0.192 & 0.021 & -0.563 & 0.014 & -1.977 & 0.328\\
Sport & -0.724 & 0.016 & -1.001 & 0.019 & -2.333 & 0.251\\
EV & -1.837 & 0.099 & -2.237 & 0.071 & -2.764 & 0.964\\
Make FEs (omitted) \\
\hline 
\textbf{SD of Random Coefficient} & & & & & & \\
$\sigma_{\text{MSRP}}$ & 0.029 & 0.014 & 0.289 & 0.004 & 1.839 & 0.210\\
\hline
\addlinespace[0.3em]
\multicolumn{7}{l}{\textbf{Own Elasticities}}\\
\hspace{1em}Average own elasticity & -1.497 &  & -2.181 &  & -5.432 & \\
\hspace{1em}SD of own elasticity  & 0.021 &  & 0.019 &  & 0.605 & \\
\hspace{1em}Minimum own elasticity & -4.102 &  & -3.673 &  & -7.886 & \\
\hspace{1em}Maximum own elasticity & -0.465 &  & -0.875 &  & -2.437 & \\
\addlinespace[0.3em]
\multicolumn{7}{l}{\textbf{Cross Elasticities (omitted)}}\\
\midrule
\multicolumn{7}{l}{\textbf{In-sample Fit}}\\
\hspace{1em}RMSE & 0.00161 &  & 0.00046 &  & 0.00037 & \\
\addlinespace[0.3em]
\multicolumn{7}{l}{\textbf{Rolling One-year-ahead Prediction}}\\
\hspace{1em}RMSE & 0.00197 &  & 0.00368 &  & 0.00403 & \\
\bottomrule
\end{tabular}
\begin{tablenotes}
\item \textit{Note: } 
The table reports posterior means and posterior standard deviations for model parameters. Own-elasticity summaries (average, SD across products, minimum, and maximum) are computed from product--market posterior mean own elasticities. The No $\eta$ specification restricts product--market demand deviations to zero, so unobserved demand varies only through market-level components. Rolling one-year-ahead prediction uses a window of 20 years for estimation and the following year for validation; we report RMSEs averaged across rolling splits.
\end{tablenotes}
\end{threeparttable}}
\end{table}
Table~\ref{tab:application_estimation_table_at} reports results for the same three specifications as in the yogurt application. 
The mean price coefficient is $-0.417$ under no-$\eta$, $-0.807$ under shrinkage, and $-3.926$ under the diffuse prior. The corresponding standard deviation of the random price coefficient rises from 0.029 to 0.289 and 1.839. Shrinkage therefore lies between the nearly homogeneous no-$\eta$ model and the highly flexible diffuse specification. The diffuse prior also produces much larger posterior standard deviations, consistent with weak identification.


The implied own-price elasticities show the same pattern. Average elasticities
are $-1.50$ under no-$\eta$, $-2.18$ under shrinkage, and $-5.43$ under
the diffuse prior, whose left tail reaches $-7.89$.
Thus, in a setting where conventional BLP elasticities vary substantially across
instrument sets, shrinkage yields more moderate estimates than the diffuse
specification.

Fit and rolling prediction mirror the yogurt tradeoff at a high level---shrinkage improves in-sample RMSE relative to no-$\eta$ ($0.00046$ versus $0.00161$) and nearly matches the diffuse prior ($0.00037$), while one-year-ahead RMSE ranks no-$\eta$ best ($0.00197$), then shrinkage ($0.00368$), then diffuse ($0.00403$); the corresponding share-fit scatterplots are in Supplementary Material Figure~\ref{fig:share_fit_compare_at}. As in the yogurt exercise, the out-of-sample gap mainly reflects the prediction design: for the holdout year, product--market deviations are drawn from the model-implied posterior predictive distribution because which products carry nonzero shocks is unknown ex ante, so these draws add noise relative to the no-$\eta$ rule that fixes them at zero---a reasonable approximation when deviations are sparse. See Supplementary Material for the details of the rolling prediction design. 

\subsubsection{Interpreting Sparse Demand Shocks}

\begin{figure}[ht]
    \centering

    \begin{subfigure}{0.48\textwidth}
        \centering
        \includegraphics[width=\linewidth]{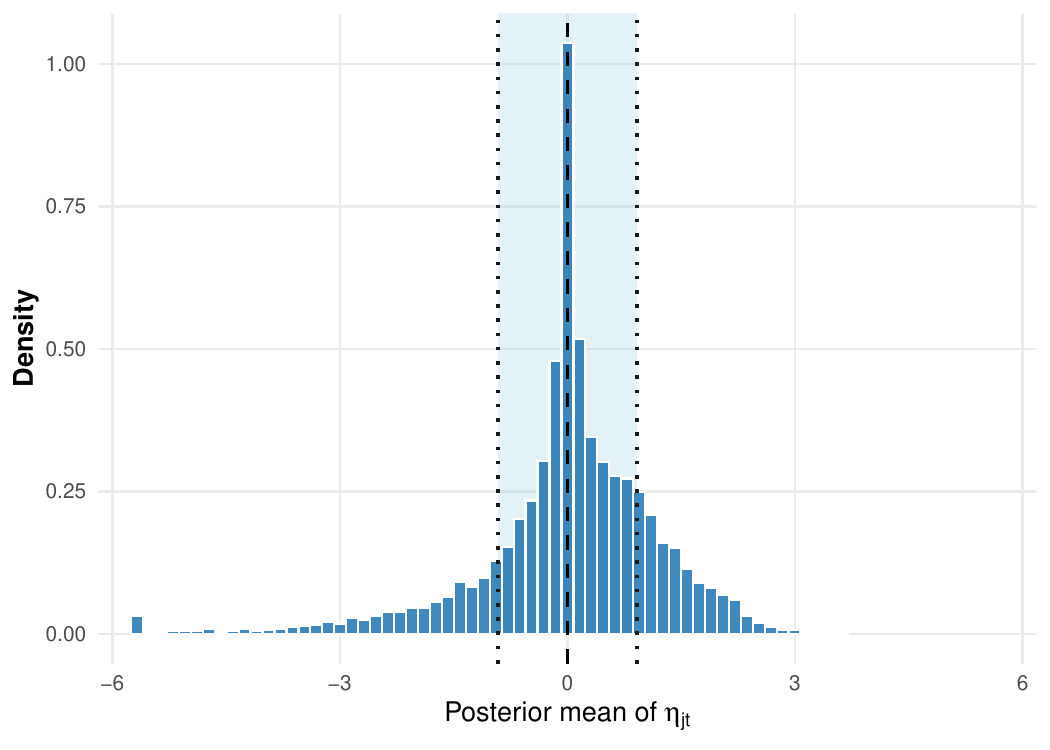}
        \caption{Shrinkage prior}
        \label{fig:at_eta_hist_shrink}
    \end{subfigure}
    \hfill
    \begin{subfigure}{0.48\textwidth}
        \centering
        \includegraphics[width=\linewidth]{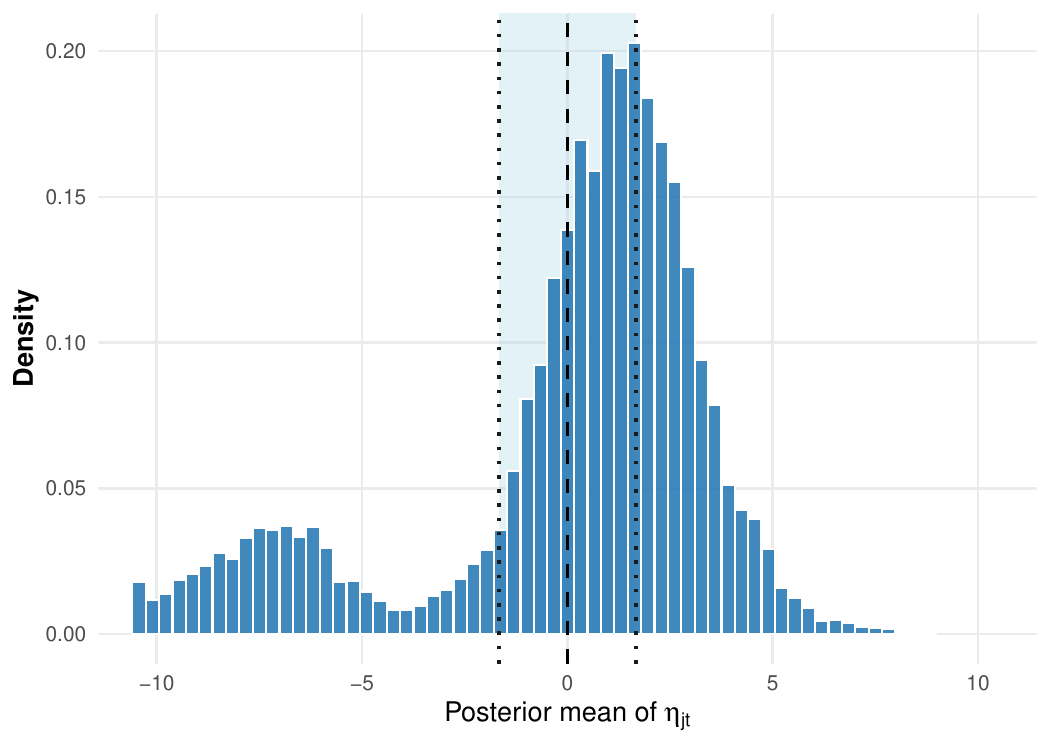}
        \caption{Diffuse prior}
        \label{fig:at_eta_hist_diffuse}
    \end{subfigure}

\caption{Distribution of posterior means of product--market deviations $\eta_{jt}$. 
The threshold $\epsilon$ is calibrated to correspond to an average 0.1 percentage-point change in predicted market share (See Section \ref{sec:calibration_threshold} for detail). 
Under the shrinkage prior (and under the diffuse prior in parentheses):
The calibrated threshold $\epsilon = 0.919$ (1.668). 
The share of $\eta_{jt}$ within $\pm \epsilon$: 67.0\% (44.2\%). 
Count near zero: 6496/9694 (4285/9694).
}
\label{fig:application_auto_eta_hist}
\end{figure}

Figure~\ref{fig:application_auto_eta_hist} illustrates a key empirical implication of the sparsity restriction. Under the shrinkage prior, posterior means of product--market deviations are sharply concentrated near zero: 67.0\% are economically negligible. Under the diffuse prior, only 44.2\% fall within the corresponding threshold, and the distribution spreads much more broadly across products and markets. The diffuse specification therefore fits shares by spreading residual variation across many \(\eta_{jt}\), which likely contributes to overfitting. By contrast, shrinkage requires most of the variation to be explained by observed covariates, random coefficients, and the market shock \(\bar\xi_t\), leaving only a sparse set of product--market deviations.

\FloatBarrier

As an ex-post check on the sparse demand shocks, Table~\ref{tab:eta_extremes_at} asks whether the product--market pairs with the largest \(|\hat\eta_{jt}|\) are economically sensible. Panel A lists the largest positive deviations. Several products recur across years---including the Nissan Leaf, Land Rover Range Rover, Ram pickup, and Volkswagen e-Golf---consistent with brand loyalty, early EV appeal, or other differentiation not fully captured by observed characteristics and make fixed effects. Panel B lists the most negative deviations: products whose shares are much weaker than their observables would predict, including several commercially weak or discontinued models (e.g., Buick Verano, Chevrolet Colorado, Lexus CT, Toyota 86).

\begin{table}[htbp]
\centering
\resizebox{.85\textwidth}{!}{
\begin{threeparttable}
\caption{Product--market pairs with the extreme demand deviations}
\label{tab:eta_extremes_at}
\begin{tabular}{rllrrrr}
\toprule
Year & Make & Model & $\hat{\eta}_{jt}$ & Post. SD & Share (\%) & MSRP \\
\midrule
\multicolumn{7}{l}{\textbf{Panel A. Largest Positive Demand Deviations}} \\
\midrule
2015 & Volkswagen & e-Golf & 3.677 & 0.237 & 0.008 & 3.627 \\
2014 & Nissan & Leaf & 3.664 & 0.217 & 0.061 & 3.289 \\
2013 & Land Rover & Range Rover & 3.569 & 0.160 & 0.025 & 9.587 \\
2004 & Land Rover & Range Rover & 3.313 & 0.246 & 0.030 & 9.066 \\
2012 & Nissan & Leaf & 3.255 & 0.264 & 0.020 & 3.827 \\
1998 & Land Rover & Range Rover & 3.197 & 0.313 & 0.017 & 8.779 \\
2017 & Land Rover & Range Rover Sport & 3.170 & 0.237 & 0.038 & 7.846 \\
2018 & Ram & Ram Pickup & 3.168 & 0.090 & 1.021 & 4.261 \\
2014 & Ram & Ram Pickup & 3.023 & 0.112 & 0.854 & 4.123 \\
2015 & Ram & Ram Pickup & 2.974 & 0.088 & 0.872 & 4.199 \\
\midrule
\multicolumn{7}{l}{\textbf{Panel B. Most Negative Demand Deviations}} \\
\midrule
2018 & Buick & Verano & -14.357 & 1.085 & 0.000 & 2.421 \\
2013 & Chevrolet & Colorado & -12.876 & 5.005 & 0.007 & 2.392 \\
2008 & BMW & Z4 & -11.511 & 7.382 & 0.013 & 4.753 \\
2015 & Lexus & CT & -11.345 & 2.514 & 0.029 & 3.098 \\
2018 & Toyota & 86 & -10.676 & 2.277 & 0.008 & 2.760 \\
2012 & Mazda & Tribute & -9.700 & 2.528 & 0.001 & 2.781 \\
2003 & Jaguar & XK & -9.665 & 1.490 & 0.007 & 9.337 \\
2014 & Chevrolet & City Express & -9.105 & 0.671 & 0.001 & 2.373 \\
2005 & Buick & Regal & -8.939 & 0.872 & 0.001 & 3.419 \\
1983 & Toyota & Land Cruiser & -8.572 & 0.689 & 0.014 & 3.181 \\
\bottomrule
\end{tabular}

\begin{tablenotes}
\footnotesize
\item \textit{Notes:} The table reports the ten product--market pairs with the largest positive and largest negative posterior mean product-level demand deviations, $\hat{\eta}_{jt}$. Positive deviations indicate products whose observed market shares substantially exceed what can be explained by observed characteristics and random-coefficient heterogeneity alone, while negative deviations indicate products whose observed shares are substantially weaker than predicted. Shares are expressed in percentage points. MSRP is measured in the same units as in the estimation data (\$10,000).
\end{tablenotes}
\end{threeparttable}
}
\end{table}
\FloatBarrier

Finally, we turn to the price endogeneity issue, which is captured by the sample correlation between price and recovered demand shocks. Because unobserved demand enters as \(\xi_{jt}=\bar\xi_t+\eta_{jt}\), sparsity does not assume away the classical endogeneity problem: prices can still correlate with the market shock \(\bar\xi_t\) and with the sparse nonzero product--market deviations. An ex-post check in Supplementary Material Table~\ref{tab:eta_msrp_regression} shows that this correlation is preserved under shrinkage: the estimated relationship between price and \(\hat\eta_{jt}\) is positive, so products with stronger residual demand tend to have higher prices, as the standard endogenous-pricing interpretation requires. We revisit this issue in the next section.

Taken together, the two applications show that sparse product--market demand shocks arise in practice and carry interpretable content, while delivering more moderate elasticities than an unrestricted treatment of \(\eta_{jt}\)---all without excluded instruments. The next section formalizes how sparsity identifies preference parameters and demand shocks from the demand system, and compares this source of identification with the conventional BLP--IV approach.

\section{Identification and Price Endogeneity}
\label{sec:identification}

\subsection{Identification under sparsity}
\label{subsec:identification_sparsity}

As discussed, the demand system (\ref{eq:demand_system}) is underidentified without further restrictions: many combinations of \((\vartheta,\xi)\) reproduce the same market shares. This subsection gives conditions under which sparsity in \(\xi\) identifies the shock vector and \(\vartheta\) from the demand system alone. For illustration, we focus on a single market and suppress the subscript \(t\) throughout. We first develop the spark argument in simple logit demand and then extend it to the random-coefficients model.

\paragraph{Simple logit.}
Under logit demand, mean utilities are
\(\delta_j=\log s_j-\log s_0=X_j^{\top}\bar\beta+\xi_j\) for
\(j=1,\ldots,J\), or \(\delta=X\bar\beta+\xi\) in matrix form. Without restrictions on \(\xi\), one can fit the observed shares for any \(\bar\beta\) by adjusting \(\xi\).

Now suppose the shock vector is sparse: \(\|\xi\|_0\le K\), where \(\|\cdot\|_0\) is the \(\ell_0\) norm, so at most \(K\)
products carry a nonzero shock.
If two \(K\)-sparse decompositions \((\bar\beta,\xi)\) and
\((\tilde{\bar\beta},\tilde\xi)\) yield the same \(\delta\), equating them gives
\(X(\bar\beta-\tilde{\bar\beta})+(\xi-\tilde\xi)=0\). The difference \(\xi-\tilde\xi\) has at most \(2K\) nonzero entries, because
\(j\) enters the difference only if \(\xi_j\neq 0\) or \(\tilde\xi_j\neq 0\).
The pair \((\bar\beta-\tilde{\bar\beta},\xi-\tilde\xi)\) therefore lies in
\(\mathrm{Null}([X\ I])\).

To state the identification condition, we follow \cite{DonohoElad2003} and define the \emph{spark} of a matrix \(M\) as the smallest support of a nontrivial null vector,
\[
\mathrm{spark}(M)\equiv\min\{\|v\|_0:\ v\neq 0,\ Mv=0\}.
\]
Writing \(v=(\bar\beta-\tilde{\bar\beta},\xi-\tilde\xi)\), we have
\(\|v\|_0\le d_X+2K\), since \(\|\xi-\tilde\xi\|_0\le 2K\) and
\(\bar\beta-\tilde{\bar\beta}\) has at most \(d_X\) entries.
If \(\mathrm{spark}([X\ I])>d_X+2K\), then any nonzero
\(v\in\mathrm{Null}([X\ I])\) has \(\|v\|_0>d_X+2K\), so the only admissible
difference is \(v=(0_{d_X},0_J)\); that is, \(\bar\beta=\tilde{\bar\beta}\) and
\(\xi=\tilde\xi\).

\begin{theorem}[Logit demand with sparse shocks]\label{thm:sparse_logit}
Let \(X\in\mathbb{R}^{J\times d_X}\) have full column rank.
Suppose \(\|\xi\|_0\le K\) and \(\mathrm{spark}([X\ I])>d_X+2K\).
Then the decomposition \(\delta=X\bar\beta+\xi\) is unique: if
\((\bar\beta,\xi)\) and \((\tilde{\bar\beta},\tilde\xi)\) both satisfy
\(\delta=X\bar\beta+\xi\) and \(\delta=X\tilde{\bar\beta}+\tilde\xi\), with
\(\|\xi\|_0\le K\) and \(\|\tilde\xi\|_0\le K\), then
\(\bar\beta=\tilde{\bar\beta}\) and \(\xi=\tilde\xi\).
\end{theorem}

\begin{remark}
Theorem~\ref{thm:sparse_logit} applies the spark-based uniqueness logic for
sparse linear representations from \cite{DonohoElad2003}.
In the single-market logit benchmark, market-level intercepts are absorbed into \(X\bar\beta\), so \(\xi\) plays the role of the product-level deviations \(\eta\) in (\ref{eq:xi}). \(K\) is the number of products with idiosyncratic demand shocks: most entries of \(\xi\) are zero, and only \(K\) products depart from the systematic component \(X\bar\beta\). Identification requires sparsity to be \emph{substantial}.
By definition, any nonzero \(v\in\mathrm{Null}([X\ I])\) satisfies
\(\|v\|_0\ge \mathrm{spark}([X\ I])\).
If two \(K\)-sparse decompositions differ, the corresponding
\(v=(\bar\beta-\tilde{\bar\beta},\xi-\tilde\xi)\) lies in \(\mathrm{Null}([X\ I])\) and satisfies \(\|v\|_0\le d_X+2K\).
Identification therefore requires that no nonzero \(v\) satisfy both bounds, i.e.\
\(\mathrm{spark}([X\ I])>d_X+2K\), or equivalently
\(K<\tfrac{1}{2}(\mathrm{spark}([X\ I])-d_X)\).\footnote{%
\(\mathrm{spark}(M)\) is also the smallest number of columns of \(M\) that are
linearly dependent \cite{DonohoElad2003}.
There is no general closed-form expression; in small problems it can be computed by searching over column subsets, though this is combinatorial in general. \cite{Donoho2006} discusses spark further in the context of sparse recovery.}
Since \(\mathrm{spark}([X\ I])\le J+d_X+1\) when the columns of \([X\ I]\) are in general position, this yields the necessary condition \(K\lesssim J/2\): at
most about half of the products can carry idiosyncratic shocks. The allowable \(K\) increases with \(\mathrm{spark}([X\ I])\): when product characteristics vary sufficiently across \(j\)---as when \(X\) has full column rank---\(\mathrm{spark}([X\ I])\) is typically large and the \(K\lesssim J/2\) bound is the relevant scale; if covariate variation is limited and \(\mathrm{spark}([X\ I])\) is small, \(K\) must be even smaller.
\end{remark}

\paragraph{Random coefficients model.}
The same spark argument extends to the random-coefficients model.
For a given \(\vartheta\), shares pin down mean utilities \(\delta(\vartheta)\)
via demand inversion, so that
\(\delta(\vartheta)=X\bar\beta+\xi\) as in the logit case.
Because \(\vartheta\) enters nonlinearly, identification is local: if two
decompositions with \(\|\xi\|_0\le K\) and \(\|\tilde\xi\|_0\le K\) yield the
same shares, linearizing around the truth gives a difference in
\(\mathrm{Null}([X\ G\ I])\) with \(\|\Delta\xi\|_0\le 2K\), where
\(G=\left.\partial\delta(\vartheta)/\partial\vartheta^{\top}\right|_{\vartheta_0}\);
the spark condition rules out nontrivial differences as above.

\begin{theorem}[Random-coefficients demand with sparse shocks]\label{thm:sparse_rc}
Let \((\bar\beta_0,\vartheta_0,\xi_0)\) satisfy
\(s_j=\sigma_j(\bar\beta_0,\xi_0,\vartheta_0)\) for all \(j=1,\ldots,J\), with
\(\|\xi_0\|_0\le K\).
Assume (i) demand inversion is locally unique and
\(\delta(\vartheta)\) is continuously differentiable near \(\vartheta_0\);
(ii) with \(G=\left.\partial\delta(\vartheta)/\partial\vartheta^{\top}\right|_{\vartheta_0}\),
\(\mathrm{spark}([X\ G\ I])>(d_X+d_\vartheta)+2K\).
Then there is a neighborhood of \((\bar\beta_0,\vartheta_0,\xi_0)\) such that any
\((\bar\beta,\vartheta,\xi)\) in that neighborhood with \(\|\xi\|_0\le K\) and
\(s_j=\sigma_j(\bar\beta,\xi,\vartheta)\) for all \(j=1,\ldots,J\) must equal
\((\bar\beta_0,\vartheta_0,\xi_0)\).
\end{theorem}

Proofs of Theorems~\ref{thm:sparse_logit} and~\ref{thm:sparse_rc} are in
Appendix~\ref{appendix:proof}.

\subsection{Comparison with BLP and Price Endogeneity}
\label{subsec:blp_endogeneity}

How does our approach compare to standard BLP? BLP achieves identification through demand inversion and excluded instruments; we do so instead by imposing sparsity on the demand shocks under (\ref{eq:xi}). Both settings involve price endogeneity, since firms observe \(\xi_{jt}\) when setting prices while econometricians do not. We first contrast these identification strategies and then discuss how sparsity shapes the correlation between price and \(\xi_{jt}\).

\paragraph{Identification via excluded IVs versus sparsity.}
In the standard BLP procedure, demand inversion pins mean utilities \(\delta_{jt}(\vartheta)\) from market shares given a candidate \(\vartheta\). The recovered demand shocks \(\xi_{jt}(\vartheta,\bar\beta)=\delta_{jt}(\vartheta)-X_{jt}^{\top}\bar\beta\) rationalize the observed shares, where \(\bar\beta\) denotes the mean taste parameters. Additional restrictions are therefore needed to identify \(\vartheta\) and \(\bar\beta\). Identification then relies on excluded instruments \(Z_{jt}\) through moment conditions \(\mathbb{E}[Z_{jt}\,\xi_{jt}(\vartheta,\bar\beta)]=0\), which hold at the true parameter values. When instruments are weak, this moment condition provides little identification information and estimates can be unstable empirically; see the sensitivity patterns in Table~\ref{tab:intro_iv_sensitivity}.  
Our approach offers an alternative route to identification: rather than excluded IV moments, it restricts the pattern of \(\xi_{jt}\) by requiring, under (\ref{eq:xi}), that most product-level deviations \(\eta_{jt}\) are zero, thereby resolving the underidentification in (\ref{eq:demand_system}).

\paragraph{Sparsity and endogeneity.}
Because firms observe \(\xi_{jt}\) when setting prices, the typical endogeneity concern is that price \(P_{jt}\) (or other endogenous characteristics in \(X_{jt}\)) is correlated with the demand shock. For illustration, consider a common linear pricing specification:
\[
P_{jt}=Z_{jt}^{\top}\rho+\omega_{jt},
\qquad
\omega_{jt}=\phi\,\xi_{jt}+\epsilon_{jt},
\]
where \(Z_{jt}\) collects exogenous shifters, \(\phi>0\), and \(\epsilon_{jt}\) is an idiosyncratic pricing residual. Under BLP, identification relies on excluded instruments \(Z_{jt}\) through orthogonality between \(\xi_{jt}\) and \(Z_{jt}\), not on restrictions on how \(\xi_{jt}\) varies across product--market pairs.

Sparsity shapes price--shock correlation in two ways. Across markets, the components \(\bar\xi_t\) are unrestricted, so sparsity places no restriction on how \(P_{jt}\) and \(\xi_{jt}\) co-vary when \(t\) changes. Within a market, by contrast, sparsity limits product-level variation in \(\xi_{jt}\): most products load on the common \(\bar\xi_t\), and only a minority can carry product-specific correlation through nonzero \(\eta_{jt}\). In the pricing specification above, substituting \(\omega_{jt}\) gives \(P_{jt}=Z_{jt}^{\top}\rho+\phi\xi_{jt}+\epsilon_{jt}\). With \(\mathbb{E}[\epsilon_{jt}\mid \xi_{jt},Z_{jt}]=0\), conditional on \(Z_{jt}\),
\[
\text{Cov}(P_{jt},\xi_{jt}\mid Z_{jt})=\phi\,\text{Var}(\xi_{jt}\mid Z_{jt}).
\]
For \(j,j'\) in the same market with \(\eta_{jt}=\eta_{j't}=0\), \(\xi_{jt}=\xi_{j't}=\bar\xi_t\), so
\[
\text{Cov}(P_{jt},\xi_{jt}\mid Z_{jt})=\text{Cov}(P_{j't},\xi_{j't}\mid Z_{j't})=\phi\,\text{Var}(\bar\xi_t\mid Z_{jt}).
\]
When \(\phi>0\), this within-market covariance is generally nonzero: products with \(\eta_{jt}=0\) remain endogenous through the shared \(\bar\xi_t\). For products with \(\eta_{jt}\neq 0\), \(\xi_{jt}=\bar\xi_t+\eta_{jt}\) has larger conditional variance, so \(\mathrm{Cov}(P_{jt},\xi_{jt}\mid Z_{jt})=\phi\,\mathrm{Var}(\xi_{jt}\mid Z_{jt})\) is correspondingly larger. Sparsity therefore does not remove endogeneity; it leaves the common market-level component unrestricted and concentrates the additional, product-specific component of price--shock correlation on the minority of active deviations. Empirically, this can be checked ex post by examining the sample association between prices and the estimated shocks \(\hat\xi_{jt}\) (or \(\hat\eta_{jt}\)), as we do in the automobile application.

\subsection{Monte Carlo Evidence}
\label{sec:simulation}

We complement the identification results in Section~\ref{subsec:identification_sparsity} and the price-endogeneity discussion in Section~\ref{subsec:blp_endogeneity} with Monte Carlo experiments in which the data-generating process satisfies (\ref{eq:xi}) with sparse product-level deviations. The goal is to assess finite-sample recovery of preference parameters and
demand shocks when excluded instruments are weak or unavailable. We compare shrinkage and diffuse-prior Bayes estimators to two BLP GMM
benchmarks: one with valid but infeasible cost instruments and one with weak instruments based only on the observed product characteristic. DGP1 keeps price exogenous; DGP2 introduces price endogeneity.

\subsubsection{Simulation Design}

We generate data from a stylized random-coefficients logit model consistent with Section~\ref{sec:model}, where the utility of consumer $i$ for product $j$ in market $t$ is
\[
u_{ijt} =  \beta_{pi} p_{jt} + \beta_w^\ast w_{jt} +  \xi_{jt}^\ast + \varepsilon_{ijt},
\]
where $\beta_{pi} \sim N(\beta_p^\ast,\sigma^{\ast 2})$ is the random coefficient on the endogenous variable price \(p_{jt}\), \(\beta_w^\ast\) is a fixed coefficient on the exogenous product characteristic $w_{jt}$,  $\varepsilon_{ijt}$ is i.i.d.\ across $i,j,t$ following the standard Gumbel distribution. 

The exogenous product characteristic $w_{jt}$ is i.i.d.\ across $j,t$ and drawn from $U(1,2)$. Price is generated as
\[
p_{jt}= \alpha_{jt}^\ast+ 0.3 w_{jt} + u_{jt},
\]
where $u_{jt}$ is a cost shock that is i.i.d.\ across $j,t$ and drawn from $N(0,.7^2)$. Unobserved demand shocks follow the within-market decomposition
\[
\xi_{jt}^\ast=\bar{\xi}_t^\ast+\eta_{jt}^\ast,
\]
where $\bar{\xi}_t^\ast=-1$ for all $t$. Within each market, 40\% of the components of $\eta_t^\ast=(\eta_{1t}^\ast,\ldots,\eta_{Jt}^\ast)^{\top}$ are nonzero and alternate in sign ($+1,-1,+1,-1,\ldots$ on the active set); the remaining 60\% are zero, following the sparse within-market structure in (\ref{eq:xi}).

The key parameters are $\beta_p^\ast=-1$, $\beta_w^\ast=0.5$, and $\sigma^\ast=1.5$. DGP1 sets $\alpha_{jt}^\ast=0$ for all $(j,t)$, so price is exogenous with respect to $\eta_t^\ast$. DGP2 lets $\alpha_{jt}^\ast$ depend on $\eta_{jt}^\ast$: $\alpha_{jt}^\ast=0.3$ if $\eta_{jt}^\ast=1$, $\alpha_{jt}^\ast=-0.3$ if $\eta_{jt}^\ast=-1$, and $\alpha_{jt}^\ast=0$ otherwise, inducing a positive correlation between $p_{jt}$ and $\xi_{jt}^\ast$.

Market shares are simulated from the random-coefficients logit using $N_t=1000$ consumer draws for the random price coefficient. Products are balanced across markets ($J_t=J$). We consider $T\in \{10,50\}$ and $J\in\{5,15\}$ and replicate each design 50 times. We implement four estimators:

\begin{itemize}[itemsep=0pt, topsep=0pt, parsep=0pt]
    \item BLP (with cost IV). An infeasible benchmark that uses $(1,w_{jt}, w_{jt}^2,  u_{jt}, u_{jt}^2)$ as instruments, where $u_{jt}$ is the unobserved cost shock in the price equation.
    
    \item BLP (without cost IV). BLP GMM with $(1,w_{jt}, w_{jt}^2, w_{jt}^3, w_{jt}^4)$ as instruments, a natural but weak choice when only $w_{jt}$ is observed.\footnote{Alternative IV sets, including sums of rivals' $w_{jt}$, performed similarly or worse.}

    \item Shrinkage. Our Bayesian estimator with the global--local shrinkage prior from Section~\ref{sec:bayes_estimation}, using $R_0=50$ normal draws to approximate choice probabilities.
    \item Diffuse. The same likelihood with a diffuse prior $\eta_{jt}\sim N(0,10^2)$, as in Section~\ref{sec:applications}.
\end{itemize}

For the Bayesian estimators, we report posterior means. For BLP, we recover $\hat{\xi}_{jt}=\hat{\delta}_{jt}- x_{jt}^{\top}\hat{\beta}$ at the estimated $\hat{\sigma}$. For each of $\beta_p,\beta_w,\sigma,$ and $\xi$, Table~\ref{tab:simulation_results} reports Monte Carlo average bias; the companion dispersion row reports Monte Carlo standard deviation for $\beta_p$, $\beta_w$, and $\sigma$, and average RMSE across product--market pairs for $\xi$.

\begin{table}[ht]
\caption{Simulation results of DGP1 and DGP2}
\label{tab:simulation_results}
\centering
\resizebox{0.95\textwidth}{!}{%
\begin{threeparttable}

\begin{subtable}{\textwidth}
\begin{tabular}{c c c|c c c c|c c c c|c c c c|c c c c}
 & &  & \multicolumn{4}{c|}{BLP (with cost IV, infeasible)} & \multicolumn{4}{c|}{BLP (without cost IV)} & \multicolumn{4}{c|}{Shrinkage} & \multicolumn{4}{c}{Diffuse} \\
$J$ & $T$ &  & $\beta_p$ & $\beta_w$ & $\sigma$ & $\xi$ & $\beta_p$ & $\beta_w$ & $\sigma$ & $\xi$ & $\beta_p$ & $\beta_w$ & $\sigma$ & $\xi$ & $\beta_p$ & $\beta_w$ & $\sigma$ & $\xi$ \\
\hline
\multirow{2}{*}{5} & \multirow{2}{*}{10} & Bias & 0.070 & -0.622 & 0.106 & 0.955 & -0.106 & -0.554 & -0.311 & 0.976 & 0.122 & -0.035 & 0.064 & 0.068 & -0.637 & -0.566 & 0.646 & 0.965 \\
 & & SD & 0.586 & 0.084 & 0.722 & 1.004 & 1.440 & 0.336 & 1.392 & 1.281 & 0.442 & 0.109 & 0.403 & 0.210 & 0.369 & 0.246 & 0.855 & 1.214 \\
\hline
\multirow{2}{*}{5} & \multirow{2}{*}{50} & Bias & 0.001 & -0.639 & 0.053 & 0.965 & 0.648 & -0.731 & -1.094 & 0.979 & -0.014 & -0.018 & 0.114 & 0.034 & -3.130 & -1.253 & 8.720 & 0.970 \\
 & & SD & 0.600 & 0.032 & 0.542 & 0.998 & 1.723 & 0.475 & 0.811 & 1.425 & 0.590 & 0.043 & 0.457 & 0.159 & 3.028 & 0.458 & 2.699 & 3.273 \\
\hline
\multirow{2}{*}{15} & \multirow{2}{*}{10} & Bias & -0.013 & -0.646 & 0.189 & 0.964 & -0.112 & -0.623 & -0.453 & 0.977 & 0.004 & -0.031 & 0.162 & 0.071 & -0.896 & -0.744 & 1.637 & 0.916 \\
 & & SD & 0.645 & 0.053 & 0.611 & 1.013 & 1.305 & 0.369 & 1.612 & 1.320 & 0.561 & 0.097 & 0.538 & 0.275 & 0.837 & 0.313 & 0.936 & 1.301 \\
\hline
\multirow{2}{*}{15} & \multirow{2}{*}{50} & Bias & -0.093 & -0.643 & 0.100 & 0.962 & 1.002 & -0.746 & -1.292 & 0.973 & -0.141 & -0.020 & 0.112 & 0.051 & -4.105 & -1.139 & 7.905 & 0.862 \\
 & & SD & 0.583 & 0.028 & 0.502 & 1.009 & 3.174 & 0.992 & 0.789 & 1.778 & 0.534 & 0.040 & 0.453 & 0.248 & 2.652 & 0.325 & 2.637 & 2.814 \\
\hline
\end{tabular}
\caption*{(a) DGP1/sparse exogenous case}
\end{subtable}

\vspace{0.5em}

\begin{subtable}{\textwidth}
\begin{tabular}{c c c|c c c c|c c c c|c c c c|c c c c}
 & &  & \multicolumn{4}{c|}{BLP (with cost IV, infeasible)} & \multicolumn{4}{c|}{BLP (without cost IV)} & \multicolumn{4}{c|}{Shrinkage} & \multicolumn{4}{c}{Diffuse} \\
$J$ & $T$ &  & $\beta_p$ & $\beta_w$ & $\sigma$ & $\xi$ & $\beta_p$ & $\beta_w$ & $\sigma$ & $\xi$ & $\beta_p$ & $\beta_w$ & $\sigma$ & $\xi$ & $\beta_p$ & $\beta_w$ & $\sigma$ & $\xi$ \\
\hline
\multirow{2}{*}{5} & \multirow{2}{*}{10} & Bias & 0.027 & -0.635 & 0.135 & 0.958 & 0.509 & -0.755 & -0.191 & 0.977 & 0.062 & -0.038 & 0.059 & 0.044 & -0.571 & -0.642 & 0.919 & 0.946 \\
 & & SD & 0.545 & 0.058 & 0.766 & 0.996 & 1.669 & 0.552 & 1.472 & 1.291 & 0.466 & 0.104 & 0.427 & 0.209 & 0.445 & 0.289 & 0.954 & 1.189 \\
\hline
\multirow{2}{*}{5} & \multirow{2}{*}{50} & Bias & -0.011 & -0.635 & 0.162 & 0.960 & -0.004 & -0.495 & -0.939 & 0.976 & 0.019 & -0.031 & 0.173 & 0.042 & -2.556 & -1.402 & 10.014 & 0.929 \\
 & & SD & 0.508 & 0.030 & 0.571 & 0.992 & 2.096 & 0.611 & 1.176 & 1.643 & 0.479 & 0.035 & 0.510 & 0.157 & 2.535 & 0.442 & 2.932 & 3.327 \\
\hline
\multirow{2}{*}{15} & \multirow{2}{*}{10} & Bias & -0.053 & -0.624 & 0.053 & 0.958 & 0.556 & -0.658 & -0.984 & 0.976 & -0.028 & -0.021 & 0.199 & 0.027 & -0.728 & -0.820 & 1.950 & 0.946 \\
 & & SD & 0.496 & 0.049 & 0.462 & 1.007 & 1.264 & 0.359 & 0.920 & 1.359 & 0.481 & 0.082 & 0.483 & 0.260 & 0.847 & 0.278 & 0.814 & 1.355 \\
\hline
\multirow{2}{*}{15} & \multirow{2}{*}{50} & Bias & 0.099 & -0.636 & 0.120 & 0.960 & 0.506 & -0.602 & -1.441 & 0.977 & 0.096 & -0.034 & 0.130 & 0.047 & -2.405 & -1.214 & 7.900 & 0.888 \\
 & & SD & 0.614 & 0.033 & 0.413 & 1.011 & 1.953 & 0.596 & 0.336 & 1.614 & 0.605 & 0.075 & 0.471 & 0.255 & 2.762 & 0.475 & 2.129 & 2.786 \\
\hline
\end{tabular}
\caption*{(b) DGP2/sparse endogenous case}
\end{subtable}

\begin{tablenotes}
\item \footnotesize{\textit{Notes:} For $\beta_p$, $\beta_w$, and $\sigma$, the Bias rows report Monte Carlo average bias and the SD rows report Monte Carlo standard deviations of the point estimates. For $\xi$, the Bias rows report the average bias of estimated demand shocks and the SD rows report the average RMSE of estimated demand shocks across product--market pairs. BLP (with cost IV) uses $(1,w_{jt},w_{jt}^2,u_{jt},u_{jt}^2)$ as instruments. BLP (without cost IV) uses $(1,w_{jt},w_{jt}^2,w_{jt}^3,w_{jt}^4)$ as instruments. Shrinkage corresponds to \texttt{rclogit\_par.stan}; Diffuse corresponds to \texttt{rclogit\_par\_normal.stan}.
50 repeated experiments. 
}
\end{tablenotes}
\end{threeparttable}%
}
\end{table}

\subsubsection{Results}

Table~\ref{tab:simulation_results} reports Monte Carlo performance under DGP1 (panel~(a)) and DGP2 (panel~(b)). BLP with cost IV is a useful but infeasible benchmark: it performs reasonably well for $\beta_p$ and $\sigma$ in many cells, but not for $\xi$. Shrinkage is the most stable estimator overall. It recovers demand shocks  more accurately than the  BLP specifications or the diffuse prior, and in most cells matches or improves on the infeasible cost-IV benchmark for the estimation of $\beta_p$, $\beta_w$, and $\sigma$. Results are similar under DGP2, so sparsity-based estimation remains stable when price is endogenous.

BLP without cost IV often exhibits large bias and dispersion in $\beta_p$ and $\sigma$, especially as $J$ and $T$ increase, consistent with the IV-sensitivity patterns in Table~\ref{tab:intro_iv_sensitivity}. The diffuse prior shows a similar identification failure in the simulations: when $J$ and $T$ are large, it absorbs share variation through many $\eta_{jt}$ (overfit), yielding large bias in $\beta_p$ and $\sigma$ and poor shock recovery. Supplementary Material Section~\ref{sec:appendix_simulation}
reports results for DGP3 and DGP4, which replace exact sparsity with dense $\eta_{jt}^\ast$; shrinkage remains comparatively stable there as well, with small bias in the structural parameters and substantially lower RMSE for latent demand shocks than other methods. Taken together, the simulations suggest that the shrinkage approach is a useful alternative when excluded instruments are weak or unavailable.

\section{Conclusion}\label{sec:conclusion}

This paper studies demand estimation for differentiated products when market--product shocks are sparse. We treat the shocks as parameters in the aggregate random-coefficients logit
likelihood, estimate them jointly with preference parameters, and impose sparsity through a within-market decomposition $\xi_{jt}=\bar\xi_t+\eta_{jt}$ together with a Bayesian shrinkage prior. Relative to standard BLP, the approach does not rely on excluded instruments or
demand inversion at each estimation step, and it delivers posterior inference for counterfactual objects such as price elasticities by integrating over uncertainty in both $\vartheta$ and the shock vector.

Our identification analysis shows how sparsity can address the
underidentification problem that arises when preference parameters and market--product demand shocks outnumber the aggregate share equations that the data provide. Under sparsity, most product-level deviations $\eta_{jt}$ are
zero within a market, so only a small number of active shocks need to be recovered alongside $\vartheta$. We formalize this logic using spark conditions: under regularity, the shock
vector and preference parameters are locally identified from market shares alone. Section~\ref{subsec:blp_endogeneity} contrasts this source of identification with the conventional BLP--IV strategy and clarifies how sparsity shapes
price--shock correlation when prices are endogenous. Bayesian shrinkage provides a practical estimator that implements the sparsity restriction while remaining scalable in high-dimensional product--market panels.

Monte Carlo experiments support the finite-sample relevance of this approach. When excluded instruments are weak or unavailable, shrinkage recovers demand shocks and structural parameters more accurately than BLP GMM and substantially
outperforms a diffuse-prior specification that allows dense product-level deviations.
Shrinkage also compares favorably to an infeasible BLP benchmark with cost instruments on shock recovery, while remaining stable under endogenous pricing and under Supplementary Material designs with dense $\eta_{jt}^\ast$.

Two empirical applications show that sparse demand shocks arise in practice and carry interpretable economic content. In store-level yogurt data, shrinkage concentrates product-level deviations on a small subset of observations, aligns with omitted promotional variables, and
produces stable elasticities and out-of-sample predictions relative to a diffuse benchmark.
In the automobile market, shrinkage recovers interpretable product--year outliers while avoiding the extreme price sensitivity and weak rolling
forecasts associated with unrestricted latent demand flexibility in a setting
where conventional BLP estimates are sensitive to instrument choice.
Together, these results suggest that sparsity is not only a technical identifying restriction but also a useful description of how unobserved demand variation is distributed across products and markets.

\begin{spacing}{1.15}
\bibliographystyle{apacite}
\bibliography{references_main}
\end{spacing}

\appendix
\clearpage   

\noindent{\huge\bfseries Appendix}
\vspace{1em}

\section{Identification Proofs}\label{appendix:proof}

This appendix proves Theorems~\ref{thm:sparse_logit}
and~\ref{thm:sparse_rc} from Section~\ref{subsec:identification_sparsity}. The argument is the same in both cases: if two sparse parameter values rationalize the same market shares, their difference lies in the null space of a stacked design matrix; sparsity limits how many coordinates can be active,
and a spark condition (\citealp{DonohoElad2003}) rules out any nontrivial difference. The logit case is linear and global; the random-coefficients case is local
because \(\vartheta\) enters the share system nonlinearly and we linearize the demand inversion map around the truth.

We work in a single market with \(J\) inside goods and suppress the market subscript~\(t\).
Let \(X\in\mathbb{R}^{J\times d_X}\) denote product characteristics,
\(\xi\in\mathbb{R}^{J}\) the demand shock vector, and
\(\delta\in\mathbb{R}^{J}\) mean utilities.
Write \(\|\cdot\|_0\) for the number of nonzero entries.
The \emph{spark} of a matrix \(M\) is
\(\mathrm{spark}(M)\equiv\min\{\|v\|_0:\ v\neq 0,\ Mv=0\}\).

\subsection{Proof of Theorem~\ref{thm:sparse_logit} (logit demand)}

Suppose \((\bar\beta,\xi)\) and \((\tilde{\bar\beta},\tilde\xi)\) satisfy
\(\delta=X\bar\beta+\xi=X\tilde{\bar\beta}+\tilde\xi\), with
\(\|\xi\|_0\le K\) and \(\|\tilde\xi\|_0\le K\).
Subtracting gives
\[
X(\bar\beta-\tilde{\bar\beta})+(\xi-\tilde\xi)=0,
\]
so \(v=(\bar\beta-\tilde{\bar\beta},\xi-\tilde\xi)\in\mathrm{Null}([X\ I])\).
Because \(\xi-\tilde\xi\) has at most \(2K\) nonzero entries,
\(\|v\|_0\le d_X+2K\).
If \(\mathrm{spark}([X\ I])>d_X+2K\), any nonzero
\(v\in\mathrm{Null}([X\ I])\) satisfies
\(\|v\|_0\ge\mathrm{spark}([X\ I])>d_X+2K\), so \(v=0\).
Hence \(\bar\beta=\tilde{\bar\beta}\) and \(\xi=\tilde\xi\).
\hfill$\square$

\subsection{Proof of Theorem~\ref{thm:sparse_rc} (random coefficients)}

Fix observed shares \(s=(s_1,\ldots,s_J)\).
For \(\vartheta\) near \(\vartheta_0\), assumption~(i) gives a
differentiable map \(\vartheta\mapsto\delta(\vartheta)\)---the demand inversion
of~\(s\) at~\(\vartheta\)---that is locally unique.
Let
\[
G=\left.\frac{\partial\delta(\vartheta)}{\partial\vartheta^{\top}}\right|_{\vartheta_0}
\in\mathbb{R}^{J\times d_\vartheta}.
\]

Suppose \((\bar\beta,\vartheta,\xi)\) and \((\bar\beta_0,\vartheta_0,\xi_0)\)
lie in a neighborhood of one another, satisfy
\(s_j=\sigma_j(\bar\beta,\xi,\vartheta)=\sigma_j(\bar\beta_0,\xi_0,\vartheta_0)\) for all~\(j\), and
obey \(\|\xi\|_0\le K\) and \(\|\xi_0\|_0\le K\).
Each parameter value rationalizes the same shares, so
\[
\delta(\vartheta)=X\bar\beta+\xi,
\qquad
\delta(\vartheta_0)=X\bar\beta_0+\xi_0,
\]
where \(\delta(\vartheta)\) and \(\delta(\vartheta_0)\) are the unique mean
utilities from demand inversion at \(\vartheta\) and \(\vartheta_0\).
When \(\vartheta\neq\vartheta_0\), \(\delta(\vartheta)\) and
\(\delta(\vartheta_0)\) need not coincide; we linearize.

Suppose \((\bar\beta,\vartheta,\xi)\neq(\bar\beta_0,\vartheta_0,\xi_0)\).
By differentiability,
\[
\delta(\vartheta)-\delta(\vartheta_0)
=G(\vartheta-\vartheta_0)+r(\vartheta,\vartheta_0),
\qquad
\|r(\vartheta,\vartheta_0)\|=o(\|\vartheta-\vartheta_0\|).
\]
Subtracting the mean-utility equations,
\[
X\Delta\bar\beta+\Delta\xi=G\Delta\vartheta+r(\vartheta,\vartheta_0),
\]
where \(\Delta\bar\beta=\bar\beta-\bar\beta_0\),
\(\Delta\vartheta=\vartheta-\vartheta_0\), and \(\Delta\xi=\xi-\xi_0\).
Equivalently, with \(w=(\Delta\bar\beta,-\Delta\vartheta,\Delta\xi)\),
\[
Xw_{\bar\beta}+Gw_{\vartheta}+w_{\xi}=r(\vartheta,\vartheta_0).
\]
Because \(\|\xi\|_0\le K\) and \(\|\xi_0\|_0\le K\), we have
\(\|w\|_0\le d_X+d_\vartheta+2K\).

Now consider any sequence of distinct parameter values in the neighborhood
that rationalize the same shares and converge to
\((\bar\beta_0,\vartheta_0,\xi_0)\). Write \(w^{(k)}\) for the associated
difference vectors and \(\hat w^{(k)}=w^{(k)}/\|w^{(k)}\|\).
Along a convergent subsequence with limit \(w^{\ast}\), we have
\(\|w^{\ast}\|_0\le d_X+d_\vartheta+2K\) and
\(\|r^{(k)}\|/\|w^{(k)}\|\to 0\) because
\(\|r\|=o(\|\Delta\vartheta\|)\) and \(\|\Delta\vartheta\|\le\|w\|\).
Passing to the limit yields \(w^{\ast}\in\mathrm{Null}([X\ G\ I])\).
Assumption~(ii) requires \(\mathrm{spark}([X\ G\ I])>(d_X+d_\vartheta)+2K\), so
any nonzero \(v\in\mathrm{Null}([X\ G\ I])\) satisfies
\(\|v\|_0\ge\mathrm{spark}([X\ G\ I])>(d_X+d_\vartheta)+2K\).
This contradicts \(\|w^{\ast}\|=1\) and \(\|w^{\ast}\|_0\le d_X+d_\vartheta+2K\).
Hence no such sequence exists, and
\((\bar\beta,\vartheta,\xi)=(\bar\beta_0,\vartheta_0,\xi_0)\).
\hfill$\square$

\section{Calibration of the Sparsity Threshold}\label{sec:calibration_threshold}
Below we describe the empirical calibration of the threshold $\epsilon>0$ used in Section~\ref{sec:applications} to summarize sparsity. We classify $\eta_{jt}$ as economically negligible or near-zero if $\vert \hat{\eta}_{jt} \vert < \epsilon$, where  $\hat{\eta}_{jt}$ is the posterior mean. 
Note that the threshold is an economically motivated reporting device, not a formal definition of sparsity in the identification argument.

To map utility perturbations into economically meaningful magnitudes, we calibrate $\epsilon$ based on the implied change in predicted market shares. 
 For each market $t$, we compute mean utilities
$
\hat\delta_{jt} = x_{jt}'\hat\beta + \hat\xi_{jt},
$
using the posterior means $\hat{\beta}$ and $\hat\xi_{jt}=\hat{\bar{\xi}}_t +\hat{\eta}_{jt}$, 
and corresponding predicted shares $\hat\sigma_{jt}$ using the model.

We then select a representative product $j(t)$ in each market, defined as the product whose predicted share is closest to the median share within that market. This choice avoids focusing on extreme products with either very high or very low baseline shares.

For a candidate perturbation $\epsilon > 0$, we compute the counterfactual shares obtained by increasing the utility of the representative product:
\[
\tilde\delta_{jt}(\epsilon) =
\begin{cases}
\hat\delta_{jt} + \epsilon & \text{if } j = j(t), \\
\hat\delta_{jt} & \text{otherwise},
\end{cases}
\]
and denote the resulting predicted shares by $\tilde\sigma_{jt}(\epsilon)$.

We then evaluate the average absolute change in own-share across markets:
\[
\Delta(\epsilon) = \frac{1}{T} \sum_{t=1}^T \left| \tilde\sigma_{j(t)t}(\epsilon) - \hat\sigma_{j(t)t} \right|.
\]

The threshold $\epsilon$ is chosen as the solution to
$
\Delta(\epsilon) = 0.001,
$
corresponding to an average change of 0.1 percentage point in predicted market share.

This procedure maps a utility perturbation into an economically interpretable change in predicted demand, ensuring that the threshold for ``near zero'' reflects economically negligible variation rather than an arbitrary numerical cutoff.

\clearpage
\part*{Supplementary Material}
\setcounter{section}{0}
\renewcommand{\thesection}{S.\arabic{section}}
\renewcommand{\thesubsection}{S.\arabic{section}.\arabic{subsection}}
\renewcommand{\thefigure}{S.\arabic{figure}}
\renewcommand{\thetable}{S.\arabic{table}}
\setcounter{figure}{0}
\setcounter{table}{0}

The supplementary material provides additional computational details, empirical
results, and simulation evidence referenced in the main text.
\section{Additional Computational Details}
\subsection{Posterior Implementation}
We simulate from the posterior distribution of the model parameters using \texttt{stan}, which implements Hamiltonian Monte Carlo. 

\paragraph{Dimensions}
\begin{itemize}[itemsep=0pt, topsep=0pt, parsep=0pt]
    \item $T$ is the number of markets.
    \item Let $\text{Jall}=\{ J_1,\ldots,J_T \}$, where $J_t$ is the number of inside options for market $t$    
    \item $TJ\equiv\sum_{t=1}^T J_t$.
    \item $d_x$ the number of covariates (the first being price).
    \item $\text{RCid}$, a $d_x$-dimensional array. 1 if the  covariate is assigned a random slope 0 
    otherwise. 
    \item $N_{rc}$ is the number of covariates with random effects, i.e. the sum of $\text{RCid}$.
    \item $R_0$ the number of $d_x$-dimensional iid $N(0,1)$ draws. 
\end{itemize}

\paragraph{Data}
The data includes, for each market $t$,
\begin{itemize}[itemsep=0pt, topsep=0pt, parsep=0pt]
    \item The  quantities sold for inside options $q_{1t},\ldots,q_{J_tt}$. 
    Let $q=(q_1',\ldots,q_T')'\in\mathbb{R}^{TJ}$ where $q_t\in\mathbb{R}^{J_t}$. 
    \item The  quantities for the outside option $q^0_{t}$. 
    Let $q^0\in\mathbb{R}^T$.
    \item $X_{jt}\in\mathbb{R}^K$, a vector of covariates. The first element is assumed to be price.\\
    Let $X_t$ be the $J_t\times d_x$ matrix. \\Let $X=(X_1',\ldots,X_T')'$ be the $TJ\times d_x$ matrix.
    \item $v$ a $d_x\times R_0$ matrix of iid $N(0,1)$ draws. 
\end{itemize}
\paragraph{Parameters}
We have 
$
\xi_{jt}=\bar{\xi}_t + \eta_{jt},
$
where $\bar{\xi}_t$ is the market-level effect and $\eta_{jt}$ is the mean-zero  deviations. 
In order to improve sampling efficiency while imposing the normalization, we work with un-restricted counterpart $\eta_{jt}^{raw}$.
\begin{itemize}[itemsep=0pt, topsep=0pt, parsep=0pt]
    \item $\beta\in\mathbb{R}^K$, a vector of slopes
    \item $\sigma_\beta \in \mathbb{R}_{+}^{N_{rc}}$, the sd for the random slopes 
    \item $\bar{\xi}=(\bar{\xi}_1,\ldots,\bar{\xi}_T)'\in\mathbb{R}^T$, the vector of market-specific effect
    \item $\eta^{raw}=(\eta_1^{raw'},\ldots,\eta_T^{raw'})'\in\mathbb{R}^{TJ}$, where $\eta_t^{raw} \in \mathbb{R}^{J_t}$
\end{itemize}
\paragraph{The Transformed Parameters}
\begin{itemize}[itemsep=0pt, topsep=0pt, parsep=0pt]
    \item $\eta=(\eta_1',\ldots,\eta_T')'\in\mathbb{R}^{TJ}$, where $\eta_t \in \mathbb{R}^{J_t}$ and 
    $\eta_{jt}=\eta_{jt}^{raw}-\frac{1}{J_t}\sum_{j=1}^{J_t}\eta_{jt}^{raw}$. This fixes the normalization that $\eta_{jt}$ is $(jt)$-specific deviation from $\bar{\xi}_t$. 

    \item $\xi=(\xi_1',\ldots,\xi_T')'\in\mathbb{R}^{TJ}$, where $\xi_t \in \mathbb{R}^{J_t}$ and $\xi_{jt}=\bar{\xi}_t + \eta_{jt}$
    \item We define $\sigma_\beta^{\text{padded}}\in\mathbb{R}_{+}^{d_x}$ as a vector of zeros, but replacing the $k$th element with the relevant term in $\sigma_\beta$ if $\text{RCid}[k]=1$. Furthermore, define a $d_x\times d_x$ matrix $R=\text{diag}(\sigma_\beta^{\text{padded}})$.
\end{itemize}

We pre-compute the $d_x\times R_0$ matrix
$
\text{RV} = R v .
$
For each market $t$, the $J_t$-dimensional vector is
\[
\delta_t = X_t \beta + \xi_t 
\]
and the $J_t\times R_0$ matrix as 
\[
\mu_t = X_t \text{RV} .
\]
We compute the shares $\sigma_t(\theta)=(\sigma_{1}(\theta),\ldots,\sigma_{J_t}(\theta))\in \mathbb{R}^{J_t}$ by first applying a column-wise softmax 
on the $J_t\times R_0$ matrix obtained by summing $\delta_t$ and $\mu_t$ 
and then taking row-wise mean.

\subsection{Computing Price Elasticities} \label{sec:compute_elasticities}
The demand elasticity of product $j$ with respect to the price change in product $m$ in market $t$ is 
\begin{equation}
E_{jm,t}=
\frac{\% \Delta \sigma_{jt}}{\% \Delta p_{mt}}
=
\frac{p_{mt}}{\sigma_{jt}}
\cdot
\frac{\partial \sigma_{jt}}{\partial p_{mt}},
\end{equation}
where 
$p_{jt}$ is the observed price of product $j$ in market $t$, 
$\beta_{price}$ is the slope on price, and 
$\sigma_{jt}$ is the model-predicted market share.
The derivative term can be written as 
\[
\frac{\partial \sigma_{jt}}{\partial p_{mt}}
=
\int 
\beta_{price,i}
\cdot
\frac{\partial \sigma_{ijt}}{\partial \delta_{mt}}
f(\beta_i) d\beta_i
=
\int 
\left( \beta_{price}+\sigma_{price}v_{i,price} \right)
\cdot
\frac{\partial \sigma_{ijt}}{\partial \delta_{mt}}
\phi(v_i \vert 0,I) d v_i,
\]
where 
$\beta_{price}$ is the slope on price, 
$\sigma_{price}$ is the standard deviation on the random coefficients on price (the first element in $\sigma_\beta$), 
and 
$v_{i,price}$ is the element in the vector $v_i$ corresponding to price. 
This is approximated by 
\[
\frac{1}{R_0}\sum_{r=1}^{R_0}
\left( \beta_{price}+\sigma_{price}v_{r,price} \right)
\cdot
\frac{\partial \sigma_{rjt}}{\partial \delta_{mt}}
\]
The partial derivatives are given as
\[
\frac{\partial \sigma_{rjt}}{\partial \delta_{mt}}
=
\begin{cases}
    \sigma_{rjt} \cdot (1-\sigma_{rjt})   \text{ if } j=m\\
    -\sigma_{rjt} \cdot \sigma_{rmt}   \text{ if } j\ne m,\\    
\end{cases}
\]
where
\[
\sigma_{rjt}=
\frac{\exp\left(\delta_{jt} + \mu_{jt}(v_r) \right)}
{1+\sum_{k=1}^{J_{t}}\exp\left(\delta_{kt} + \mu_{kt}(v_r) \right)}
\]

\section{Additional Results from the Empirical Applications}

\subsection{Yogurt Market}

\begin{figure}[ht]
    \centering

    \begin{subfigure}{0.31\textwidth}
        \centering
        \includegraphics[width=\linewidth]{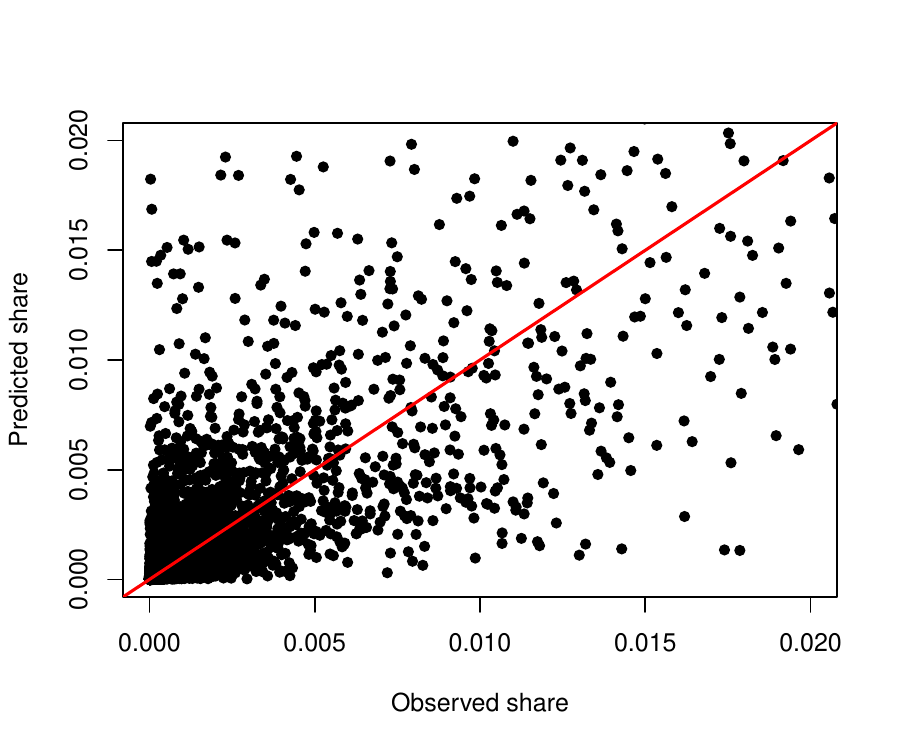}
        \caption{No $\eta$}
        \label{fig:yg_share_fit_noeta}
    \end{subfigure}
    \hfill
    \begin{subfigure}{0.31\textwidth}
        \centering
        \includegraphics[width=\linewidth]{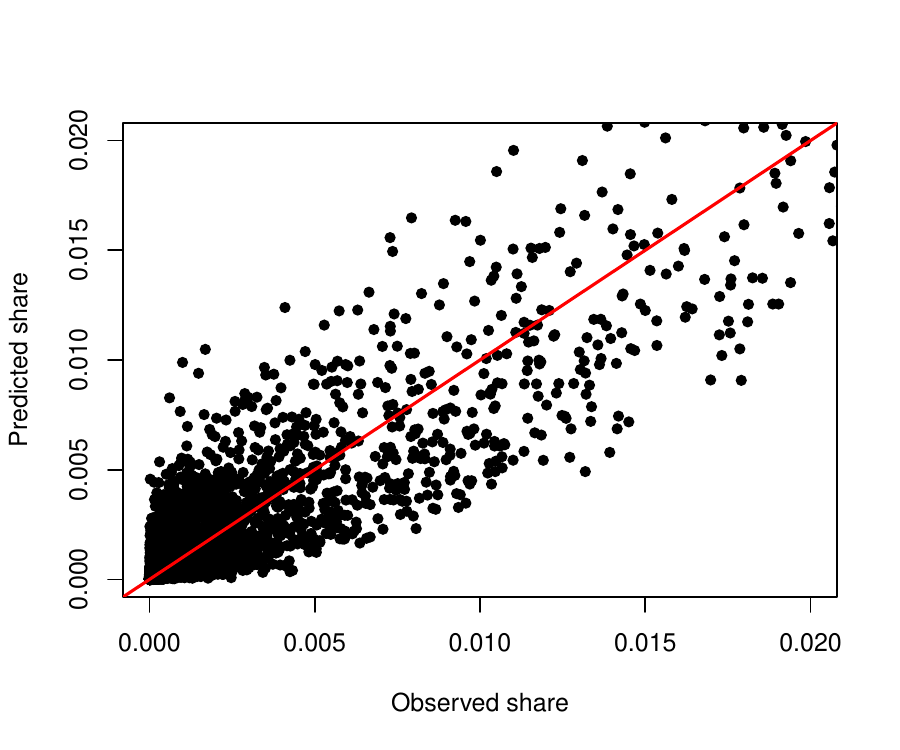}
        \caption{Shrinkage prior}
        \label{fig:yg_share_fit_shrink}
    \end{subfigure}
    \hfill
    \begin{subfigure}{0.31\textwidth}
        \centering
        \includegraphics[width=\linewidth]{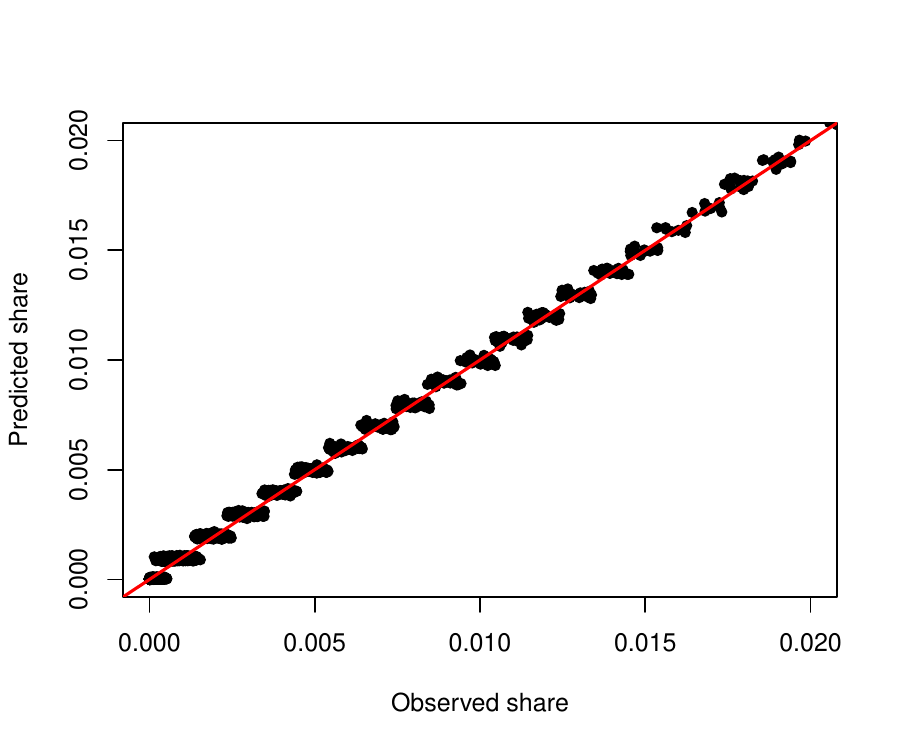}
        \caption{Diffuse prior}
        \label{fig:yg_share_fit_diffuse}
    \end{subfigure}

\caption{Observed versus predicted market shares. Each point corresponds to a product--store pair. The red 45-degree line indicates perfect fit.}
\label{fig:share_fit_compare_yg}
\end{figure}

\paragraph{In-sample share fit.}
Figure~\ref{fig:share_fit_compare_yg} plots predicted versus observed market shares for the three specifications. The no-$\eta$ specification underfits realized shares; the diffuse prior fits almost perfectly, consistent with overfitting; and shrinkage sits in between. The corresponding in-sample RMSEs are reported in Table~\ref{tab:estimation_table_yg} in the main text.
\FloatBarrier
\FloatBarrier

\paragraph{Out-of-sample prediction design.}
We evaluate out-of-sample predictive performance using five-fold store-level cross-validation. Stores are randomly partitioned into five groups. For each fold, the model is estimated on the training stores and then used to predict market shares in the held-out stores. Prediction is conducted draw by draw.

For posterior draw \(r\), we sample held-out product deviations from the corresponding posterior predictive distributions. For the shrinkage specification, we draw one new product deviation for each held-out product--store observation from the global--local prior:
$\lambda_{jt}^{\mathrm{new},(r)}\sim C^+(0,1)$
and
$\eta_{jt}^{\mathrm{new},(r)}
\sim
N\!\left(0,\left(\tau_0^{(r)}\lambda_{jt}^{\mathrm{new},(r)}\right)^2\right)$,
where \(\tau_0^{(r)}\) is the \(r\)-th posterior draw of the global shrinkage parameter. For the diffuse-prior specification, we draw
$\eta_{jt}^{\mathrm{new},(r)}
\sim
N(0,10^2)$.
For the no-\(\eta\) specification, held-out product deviations are fixed at zero,
$\eta_{jt}^{\mathrm{new},(r)}=0.$
We set the market-level demand component for a held-out store equal to the average of the market-level components estimated from the training stores,
$
\bar{\xi}_{\mathrm{new}}^{(r)}
=
\frac{1}{T_{\mathrm{train}}}
\sum_{t\in\mathcal T_{\mathrm{train}}}
\bar{\xi}_{t}^{(r)}.
$
Given these draws, the predicted share for product \(j\) in held-out store \(t\) is
$s_{jt}^{\mathrm{new},(r)}
=
s_{jt}\!\left(
\bar{\beta}^{(r)},\Sigma^{(r)},
\bar{\xi}_{\mathrm{new}}^{(r)}
+
\eta_{jt}^{\mathrm{new},(r)}
\right)$.
Predictive accuracy is measured by RMSE across held-out product--store observations; the resulting RMSEs are reported in Table~\ref{tab:estimation_table_yg}.

The ranking favors no-$\eta$ (0.0089), then shrinkage (0.0161), then the diffuse prior (0.0264). This ordering mainly reflects the prediction design rather than a failure of shrinkage in sample: held-out stores receive freshly drawn product deviations, and because which products are promoted is unknown ex ante, these draws add noise relative to the no-$\eta$ rule that fixes them at zero---a good approximation when deviations are sparse. The posterior mean of the global shrinkage parameter is \(\hat{\tau}_0=0.382\), so new deviations under shrinkage are regularized but not forced to zero; under the diffuse prior they are drawn from \(N(0,10^2)\), which is extremely dispersed on the utility scale. An alternative rule that sets $\eta_{jt}^{\mathrm{new}}=0$ for all specifications is discussed below.

\begin{figure}[!ht]
\centering

\begin{minipage}{0.49\textwidth}
\centering
\includegraphics[width=\textwidth]{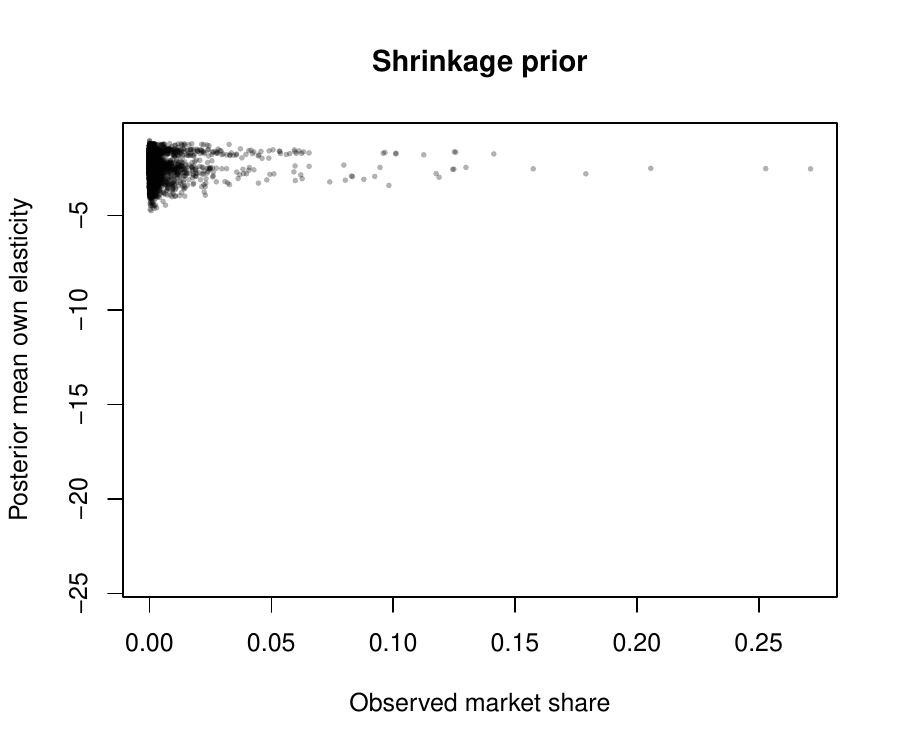}
\caption*{(a) Shrinkage prior}
\end{minipage}
\hfill
\begin{minipage}{0.49\textwidth}
\centering
\includegraphics[width=\textwidth]{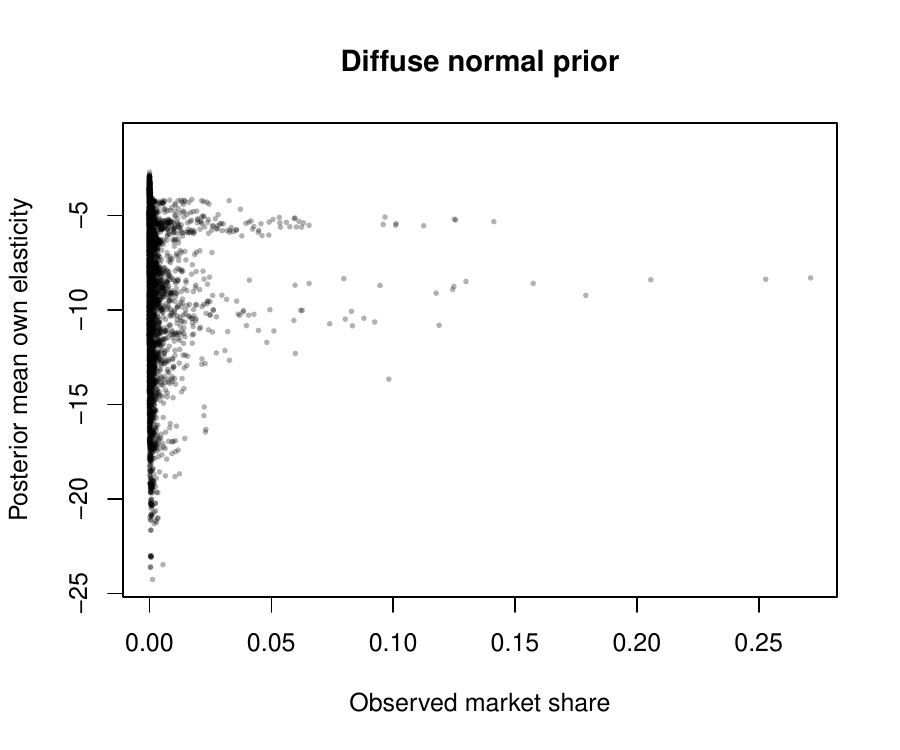}
\caption*{(b) Diffuse prior}
\end{minipage}

\caption{
Posterior mean own-price elasticities plotted against observed market shares. 
}
\label{fig:yg_elasticity_share}
\end{figure}

\paragraph{Own-price elasticities and small-share products.}
Figure~\ref{fig:yg_elasticity_share} plots posterior mean own-price elasticities against observed market shares for the shrinkage and diffuse specifications. Under the diffuse prior, the most extreme elasticities are concentrated among products with very small shares. In a simple logit model, \(\varepsilon_{jj}=\alpha p_j(1-s_j)\), so small shares alone do not mechanically imply large elasticities; for \(s_j\approx 0\), the term \(1-s_j\) is close to one. The extreme values therefore primarily reflect the large estimated price sensitivity under the diffuse specification. The no-$\eta$ and shrinkage specifications produce more moderate elasticities (Table~\ref{tab:estimation_table_yg}).

\FloatBarrier

\begin{table}[!h]
\centering
\caption{Posterior Estimates and Holdout Prediction with Zero Product Deviations}
\label{tab:estimation_table_yg_appdx}
\centering
\resizebox{0.6\linewidth}{!}{
\begin{threeparttable}
\begin{tabular}[t]{lcccccc}
\toprule
\multicolumn{1}{c}{ } & \multicolumn{2}{c}{No $\eta$} & \multicolumn{2}{c}{Shrinkage prior} & \multicolumn{2}{c}{Diffuse prior} \\
\cmidrule(l{3pt}r{3pt}){2-3} \cmidrule(l{3pt}r{3pt}){4-5} \cmidrule(l{3pt}r{3pt}){6-7}
Variable & Mean & SD & Mean & SD & Mean & SD\\
\midrule
Price & -1.120 & 0.030 & -1.474 & 0.020 & -4.640 & 0.232\\
AXELROD & -0.242 & 0.065 & -0.341 & 0.037 & -4.096 & 1.253\\
CABOT & 0.423 & 0.246 & 0.246 & 0.172 & -0.979 & 1.576\\
CHOBANI & 2.396 & 0.048 & 2.099 & 0.038 & 4.686 & 0.746\\
DANNON ALL NATURAL & -0.350 & 0.050 & -0.662 & 0.043 & -3.301 & 0.721\\
DANNON LIGHT N FIT & 0.874 & 0.042 & 1.067 & 0.052 & 1.508 & 0.863\\
DANNON OIKOS & 0.507 & 0.066 & 0.713 & 0.050 & 2.776 & 0.972\\
FAGE TOTAL & 2.622 & 0.073 & 2.126 & 0.057 & 6.061 & 0.878\\
LA YOGURT & -0.239 & 0.045 & -0.278 & 0.050 & -1.566 & 0.814\\
PRIVATE LABEL & -0.425 & 0.041 & -0.554 & 0.034 & -2.161 & 0.601\\
STONYFIELD ORGANIC & 0.423 & 0.126 & 1.091 & 0.158 & 2.548 & 1.201\\
STONYFIELD ORGANIC OIKOS & 1.893 & 0.178 & 2.702 & 0.120 & 9.562 & 1.448\\
VOSKOS & -0.284 & 0.144 & 0.031 & 0.167 & 0.846 & 1.200\\
YOPLAIT & -1.940 & 0.134 & -1.840 & 0.097 & -5.442 & 1.116\\
YOPLAIT LIGHT & -0.284 & 0.048 & 0.010 & 0.047 & -0.440 & 0.796\\
YOPLAIT ORIGINAL & 0.210 & 0.054 & 0.245 & 0.063 & 0.432 & 1.019\\
Size 2 & -3.076 & 0.116 & -2.868 & 0.098 & -10.944 & 0.959\\
Size 3 & -5.400 & 0.850 & -4.828 & 0.718 & -10.717 & 1.913\\
Size 4 & -3.062 & 0.051 & -2.915 & 0.062 & -10.574 & 0.524\\
Flavored & 1.642 & 0.035 & 1.175 & 0.034 & 0.961 & 0.446\\
Nonfat & 1.179 & 0.076 & 0.729 & 0.070 & -0.024 & 0.513\\
Lowfat & 0.730 & 0.076 & 0.483 & 0.055 & -0.727 & 0.525\\
Greek & 0.167 & 0.052 & 0.195 & 0.048 & 0.343 & 0.754\\
Organic & -0.543 & 0.130 & -0.918 & 0.155 & -3.106 & 1.041\\
\midrule
$\sigma_{\text{Price}}$ & 0.290 & 0.017 & 0.479 & 0.013 & 0.657 & 0.088\\
$\sigma_{\text{Organic}}$ & 0.517 & 0.211 & 0.804 & 0.174 & 0.937 & 0.498\\

\midrule




\multicolumn{7}{l}{\textbf{Cross Elasticities}}\\
\hspace{1em}Average cross elasticity & 0.007 &  & 0.009 &  & 0.047 & \\
\hspace{1em}SD of cross elasticity & 0.000 &  & 0.002 &  & 0.019 & \\
\hspace{1em}Minimum cross elasticity & 0.000 &  & 0.000 &  & 0.000 & \\
\hspace{1em}Maximum cross elasticity & 0.514 &  & 1.043 &  & 4.502 & \\
\midrule
\addlinespace[0.3em]
\multicolumn{7}{l}{\textbf{Out-of-sample Prediction ($\eta_{jt}^{new}=0$)}}\\
\hspace{1em}RMSE & 0.0089 &  & 0.0092 &  & 0.0095 & \\

\bottomrule
\end{tabular}
\begin{tablenotes}
\item \textit{Note: } The table reports posterior means and posterior standard deviations for model parameters. Brand 1 and Size 1 are omitted baseline categories. The specification includes random coefficients on price and organic. Cross-elasticity summaries are computed across product--store pairs. The No $\eta$ specification restricts product-level demand deviations to zero, so unobserved demand varies only through market-level components.
\item Out-of-sample prediction uses the same store-level holdout design as in the main text, except that held-out product deviations are set to $\eta_{jt}^{\mathrm{new}}=0$ for all specifications. RMSEs are computed across product--store observations in the held-out stores.
\end{tablenotes}
\end{threeparttable}}
\end{table}
\FloatBarrier

\paragraph{Holdout prediction with zero new product deviations.}
As an additional check, we evaluate held-out store performance under a more restrictive prediction rule. For all three specifications, product deviations in held-out stores are set to their posterior predictive mean of zero,
\[
\eta_{jt}^{\mathrm{new}}=0.
\]
As before, we set the market-level demand component for a held-out store equal to the average of the training-store market components, per MCMC iteration:
$
\bar{\xi}_{\mathrm{new}}^{(r)}
=
\frac{1}{T_{\mathrm{train}}}
\sum_{t\in\mathcal T_{\mathrm{train}}}
\bar{\xi}_{t}^{(r)}.
$
This exercise asks how well the estimated systematic demand component transfers to new stores when product-level shocks are unavailable at prediction time.

Table~\ref{tab:estimation_table_yg_appdx} reports the results. The no-\(\eta\) specification achieves RMSE 0.0089; shrinkage is nearly identical (0.0092); and the diffuse prior is slightly worse (0.0095). Thus, when held-out deviations are fixed at zero, shrinkage nearly matches the restrictive benchmark and improves on the diffuse prior---complementing the main-text posterior predictive exercise, where freshly drawn deviations add noise relative to no-$\eta$.

\paragraph{Regression check with recorded Display and Feature.}
The main text validates \(\hat\eta_{jt}\) using mean lifts by promotion status. Table~\ref{tab:eta_marketing_mix_regression_yg} reports the corresponding regression. Under shrinkage, both recorded \textit{Display} and \textit{Feature} enter positively. Under the diffuse prior, \textit{Display} remains positive but \textit{Feature} turns negative. Together with the much wider distribution of \(\hat\eta_{jt}\) under the diffuse prior, this suggests that diffuse product deviations are primarily absorbing realized share variation rather than recovering the same promotional pattern.

\FloatBarrier
\begin{table}[h!]
\centering
\begin{threeparttable}
\caption{Relationship between latent demand deviations and marketing mix variables}
\label{tab:eta_marketing_mix_regression_yg}

\begin{tabular}{lcc}
\toprule
 & Shrinkage prior & Diffuse prior \\
\midrule

Intercept & -0.008* & 0.018 \\
 & (0.005) & (0.057) \\

Display & 0.205*** & 1.593*** \\
 & (0.033) & (0.412) \\

Feature & 0.032** & -0.520*** \\
 & (0.014) & (0.179) \\

\midrule
$R^2$ & 0.009 & 0.003 \\
Adjusted $R^2$ & 0.009 & 0.003 \\
Observations & 5927 & 5927 \\

\bottomrule
\end{tabular}

\begin{tablenotes}
\footnotesize
\item \textit{Notes:} The dependent variable is the posterior mean of the product-specific demand deviation $\eta_{jt}$. Standard errors are reported in parentheses. $^{***}p<0.01$, $^{**}p<0.05$, $^{*}p<0.1$.
\end{tablenotes}

\end{threeparttable}
\end{table}
\FloatBarrier

\begin{figure}[!ht]
\centering

\begin{minipage}{0.49\textwidth}
\centering
\includegraphics[width=\textwidth]{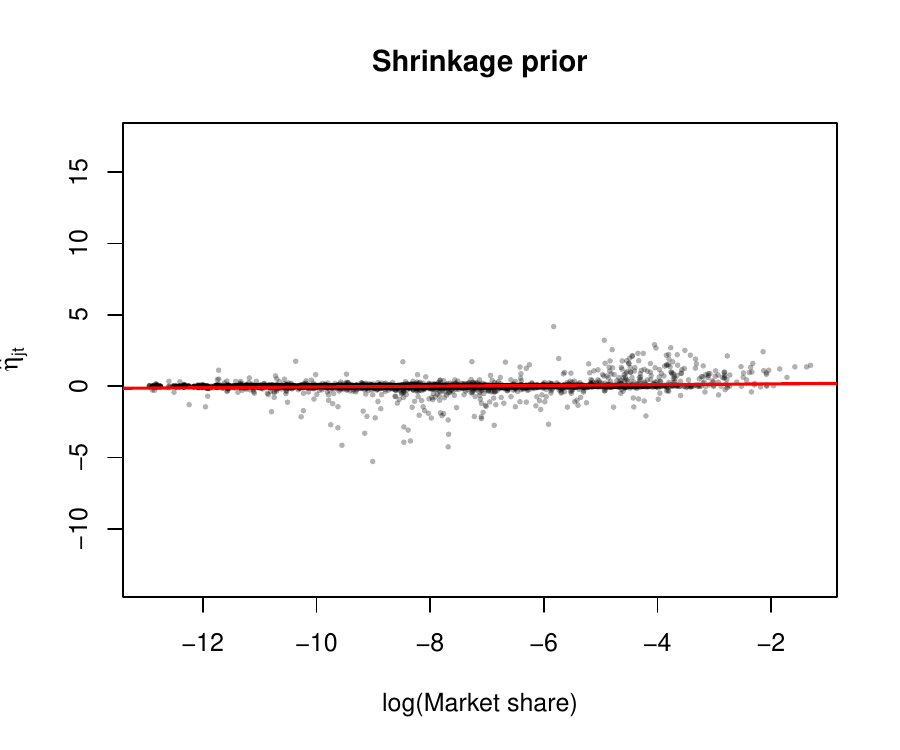}
\caption*{(a) Shrinkage prior}
\end{minipage}
\hfill
\begin{minipage}{0.49\textwidth}
\centering
\includegraphics[width=\textwidth]{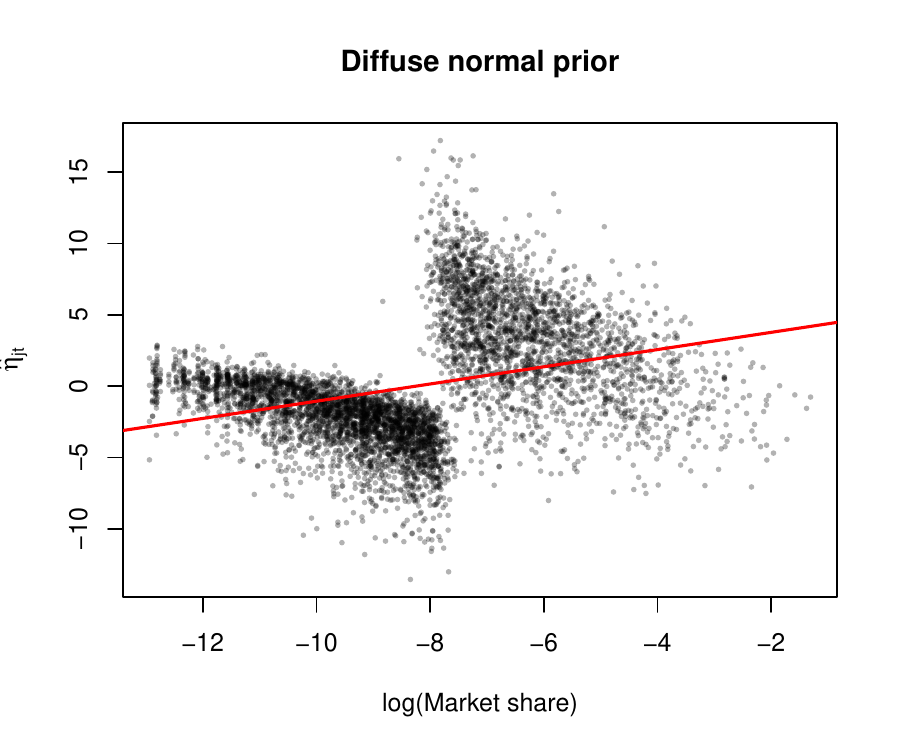}
\caption*{(b) Diffuse prior}
\end{minipage}

\caption{
Posterior mean demand deviations $\hat{\eta}_{jt}$ plotted against log market shares. The red line shows the fitted linear relationship. Under the diffuse prior, the estimated demand deviations exhibit substantially greater dispersion and a stronger relationship with observed shares, suggesting that the latent demand shocks are used more aggressively to fit store-level outcomes. The shrinkage prior produces more concentrated demand shocks and attenuates this dependence.
}
\label{fig:share_eta_comparison_yg}
\end{figure}

Figure~\ref{fig:share_eta_comparison_yg} plots the posterior means of $\eta_{jt}$ against observed shares. Under the diffuse prior, the estimated demand deviations exhibit substantially greater dispersion and a much stronger relationship with observed shares: products with very small shares tend to receive large negative deviations, while products with large shares receive large positive deviations. This pattern suggests that $\eta_{jt}$ is being used aggressively to fit realized market shares. In contrast, the shrinkage prior substantially attenuates this relationship, consistent with the view that the diffuse prior tends to overfit product--store variation, whereas shrinkage regularizes the latent demand component.

\FloatBarrier

Table~\ref{tab:eta_price_regression_yg} shows that under both priors, estimated $\eta_{jt}$ is positively associated with price, so products with stronger residual demand tend to have higher prices.
\begin{table}[h!]
\centering
\begin{threeparttable}
\caption{Relationship between latent demand deviations and price}
\label{tab:eta_price_regression_yg}

\begin{tabular}{lcc}
\toprule
 & Shrinkage prior & Diffuse prior \\
\midrule

Intercept & -0.023** & -0.732*** \\
 & (0.010) & (0.123) \\

price & 0.008** & 0.255*** \\
 & (0.003) & (0.039) \\

\midrule
$R^2$ & 0.001 & 0.007 \\

Observations & 5927 & 5927 \\

\bottomrule
\end{tabular}

\begin{tablenotes}
\footnotesize
\item \textit{Notes:} The dependent variable is the posterior mean of the product-specific demand deviation $\eta_{jt}$. Price is measured in the same units as in the estimation data. Standard errors are reported in parentheses. $^{***}p<0.01$, $^{**}p<0.05$, $^{*}p<0.1$.
\end{tablenotes}

\end{threeparttable}
\end{table}
\FloatBarrier

\subsection{Automobile Market}
\label{sec:appendix_auto}

\paragraph{Rolling one-year-ahead prediction.}
\label{sec:appendix_auto_rolling}
To compare predictive performance, we use a rolling-window design. In each window, the model is estimated on the preceding \(\tilde{T}=20\) years and then used to predict market shares in the following year. For example, one window uses 1998--2017 to predict 2018; the next uses 1997--2016 to predict 2017. This design mirrors a realistic forecasting exercise for automobile manufacturers, who must predict demand for the coming model year using only information available through the current year together with the covariates in the upcoming year. We use 10 rolling splits.

Since $\bar{\xi}_{t}$ is unobserved for the holdout year ($t=\tilde{T}+1$), we approximate it by the posterior draw of the final in-sample value, setting
$
\bar{\xi}_{\tilde{T}+1}^{(m)}=\bar{\xi}_{\tilde{T}}^{(m)},
$
for every posterior draw \(m\). Conditional on this value, predicted market shares use the posterior draws of the remaining parameters. For the shrinkage and diffuse-prior specifications, \(\eta_{j,\tilde{T}+1}\) are sampled from the corresponding posterior predictive distributions; for the no-\(\eta\) specification they are fixed at zero, as in the yogurt prediction exercise.\footnote{We obtained similar results when we set \(\eta_{j,\tilde{T}+1}=0\) for all specifications.} Prediction accuracy is summarized by the RMSE for each holdout year, then averaged across rolling splits; the resulting RMSEs are reported in Table~\ref{tab:application_estimation_table_at}.

\begin{figure}[!ht]
\centering

\begin{minipage}{0.49\textwidth}
\centering
\includegraphics[width=\textwidth]{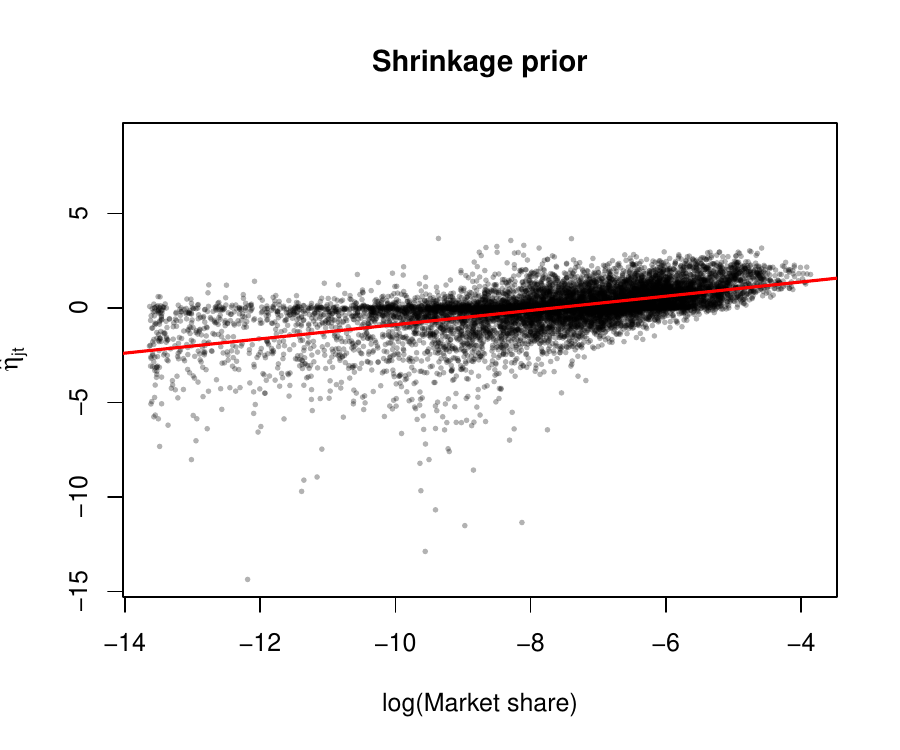}
\caption*{(a) Shrinkage prior}
\end{minipage}
\hfill
\begin{minipage}{0.49\textwidth}
\centering
\includegraphics[width=\textwidth]{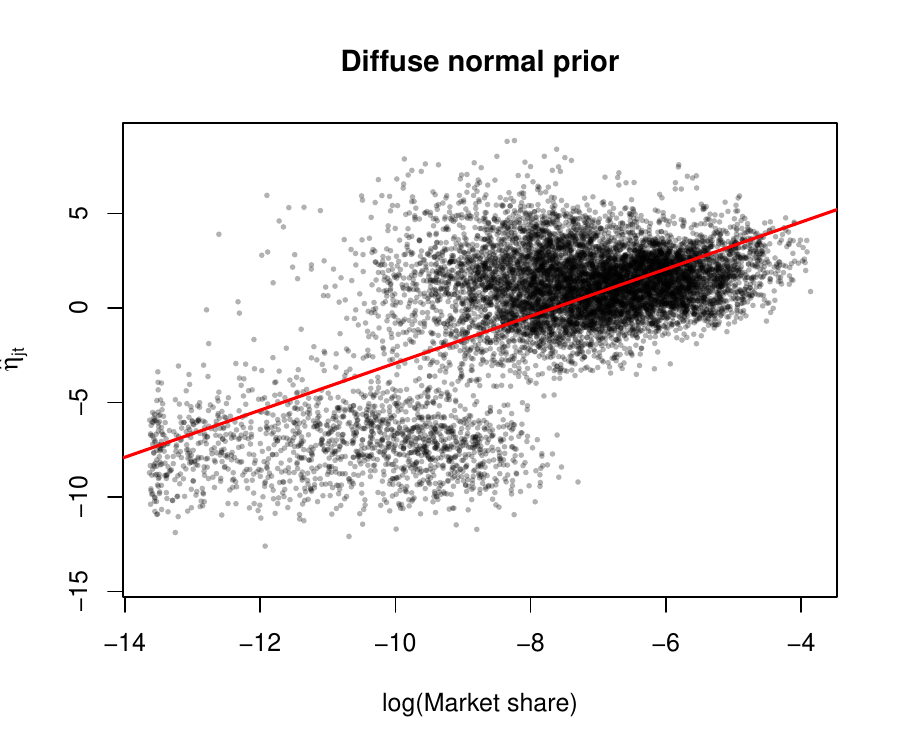}
\caption*{(b) Diffuse prior}
\end{minipage}

\caption{
Posterior mean demand deviations $\hat{\eta}_{jt}$ plotted against log market shares. The red line shows the fitted linear relationship.
}
\label{fig:share_eta_comparison_at}
\end{figure}

\paragraph{Product--market deviations and market shares.}
Figure~\ref{fig:share_eta_comparison_at} plots posterior mean product--market deviations against log observed market shares. Both specifications exhibit a positive relationship, but the diffuse prior relies much more heavily on extreme latent demand adjustments. In particular, it assigns large negative \(\eta_{jt}\)'s to products with extremely small observed shares; for example, a log share around \(-13\) corresponds to a share of approximately \(e^{-13}\simeq 10^{-6}\). This flexibility helps the diffuse specification fit observed shares in sample, but it also makes the fitted latent demand component less transferable to new markets. The rolling prediction results show the consequence: the diffuse model's superior in-sample fit does not translate into better out-of-sample prediction. The shrinkage prior regularizes these extreme adjustments, yielding a lower-dimensional latent demand structure that generalizes better while still allowing economically meaningful product--market deviations.
\FloatBarrier

\begin{figure}[ht]
    \centering

    \begin{subfigure}{0.31\textwidth}
        \centering
        \includegraphics[width=\linewidth]{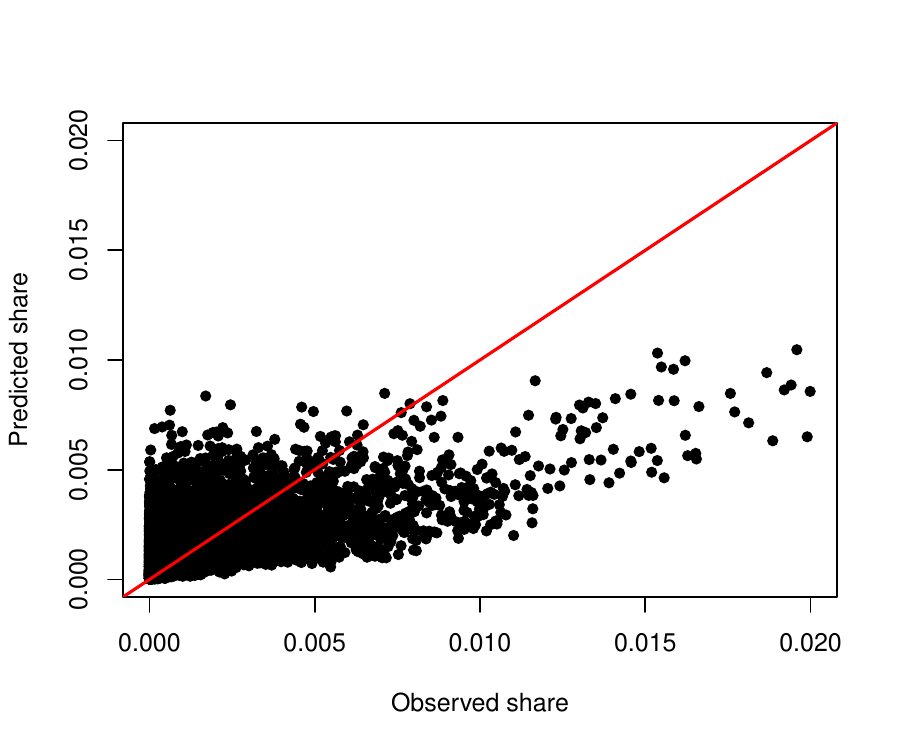}
        \caption{No $\eta$}
        \label{fig:at_share_fit_noeta}
    \end{subfigure}
    \hfill
    \begin{subfigure}{0.31\textwidth}
        \centering
        \includegraphics[width=\linewidth]{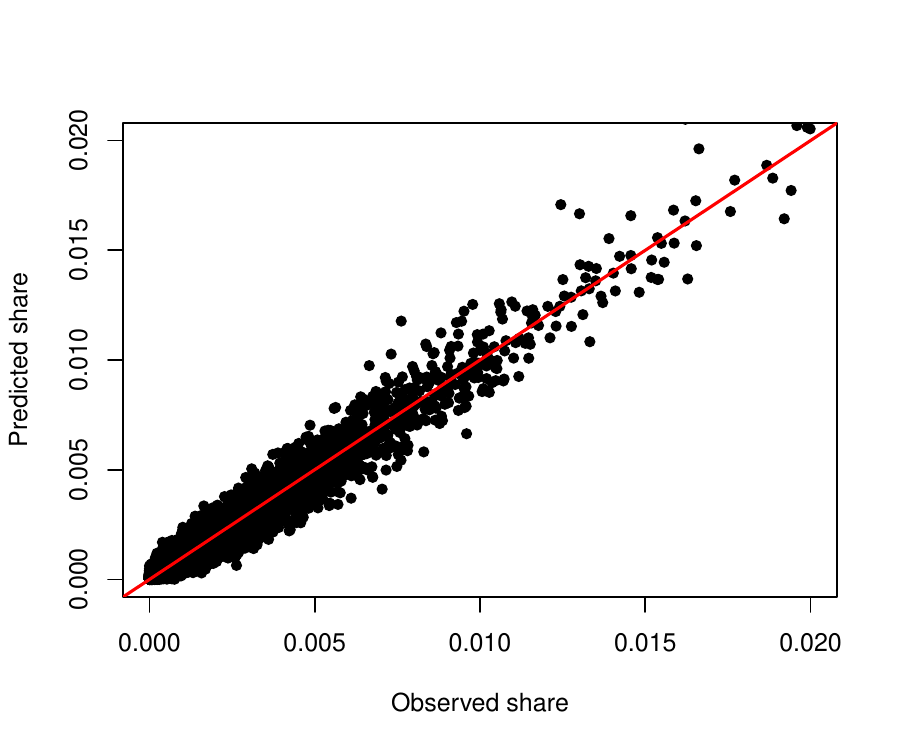}
        \caption{Shrinkage prior}
        \label{fig:at_share_fit_shrink}
    \end{subfigure}
    \hfill
    \begin{subfigure}{0.31\textwidth}
        \centering
        \includegraphics[width=\linewidth]{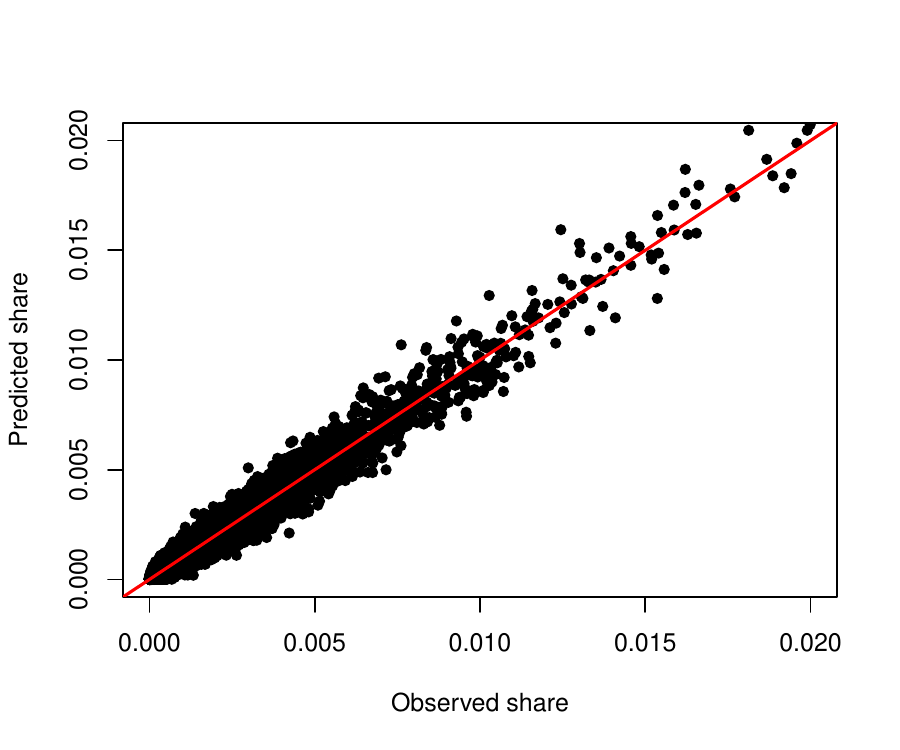}
        \caption{Diffuse prior}
        \label{fig:at_share_fit_diffuse}
    \end{subfigure}

\caption{Observed versus predicted market shares. Each point corresponds to a product--market pair. The red 45-degree line indicates perfect fit.}
\label{fig:share_fit_compare_at}
\end{figure}

\paragraph{In-sample share fit.}
Figure~\ref{fig:share_fit_compare_at} plots predicted versus observed market shares. As in the yogurt application, no-$\eta$ underfits, the diffuse prior fits almost perfectly, and shrinkage sits in between; the corresponding RMSEs are in Table~\ref{tab:application_estimation_table_at}.
\FloatBarrier

\paragraph{Price endogeneity: correlation between MSRP and $\hat\eta_{jt}$.}
Table~\ref{tab:eta_msrp_regression} reports the sample relationship between price and recovered product--market deviations. Because unobserved demand enters as \(\xi_{jt}=\bar\xi_t+\eta_{jt}\), sparsity does not assume away classical endogeneity: prices can still correlate with the market shock \(\bar\xi_t\) and with the sparse nonzero product--market deviations.

\begin{table}[h!]
\centering
\begin{threeparttable}
\caption{Relationship between latent demand deviations and MSRP}
\label{tab:eta_msrp_regression}

\begin{tabular}{lcc}
\toprule
 & Shrinkage prior & Diffuse prior \\
\midrule

Intercept & -0.133*** & 0.148* \\
 & (0.030) & (0.086) \\

MSRP & 0.037*** & -0.041* \\
 & (0.007) & (0.021) \\

\midrule
$R^2$ & 0.003 & 0.000 \\

Observations & 9694 & 9694 \\

\bottomrule
\end{tabular}

\begin{tablenotes}
\footnotesize
\item \textit{Notes:} The dependent variable is the posterior mean of the product--market demand deviation $\eta_{jt}$. MSRP (price) is measured in the same units as in the estimation data. Standard errors are reported in parentheses. $^{***}p<0.01$, $^{**}p<0.05$, $^{*}p<0.1$.
\end{tablenotes}

\end{threeparttable}
\end{table}

Under the shrinkage prior, the estimated relationship between price and latent demand deviations is positive and highly significant, consistent with the standard endogenous-pricing interpretation: products with stronger residual demand tend to command higher prices. In contrast, the diffuse-prior specification produces a weakly negative relationship with essentially no explanatory power.\footnote{The very small \(R^2\) values are not particularly concerning here: the latent demand deviations capture much richer product--market heterogeneity than price alone can explain. The regression is meant only to illustrate the ex-post correlation between price and $\eta_{jt}$.}

\begin{figure}[!ht]
\centering

\begin{minipage}{0.49\textwidth}
\centering
\includegraphics[width=\textwidth]{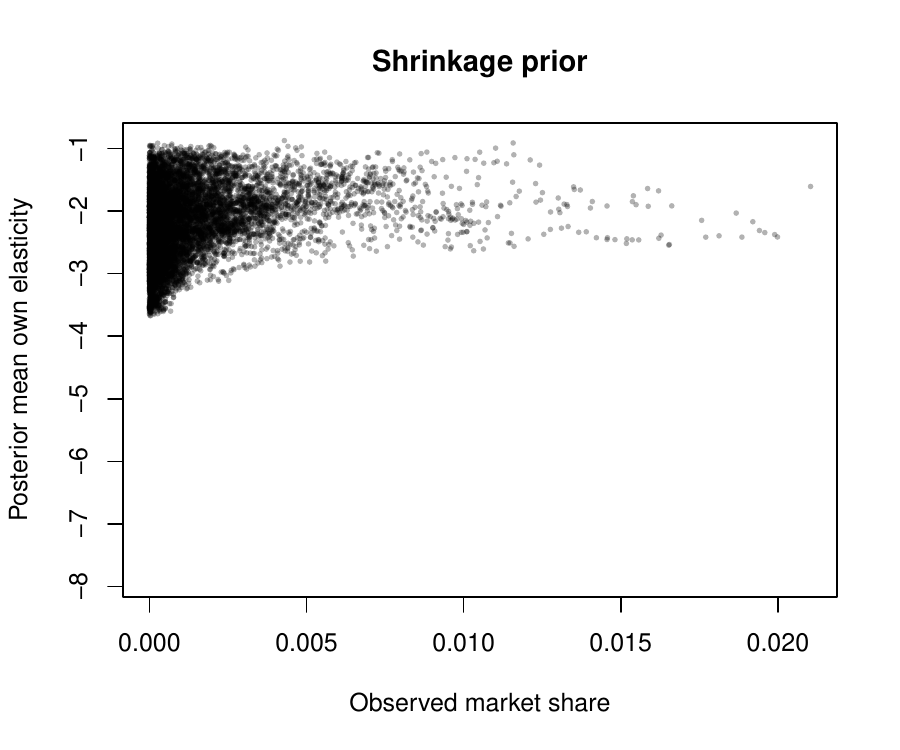}
\caption*{(a) Shrinkage prior}
\end{minipage}
\hfill
\begin{minipage}{0.49\textwidth}
\centering
\includegraphics[width=\textwidth]{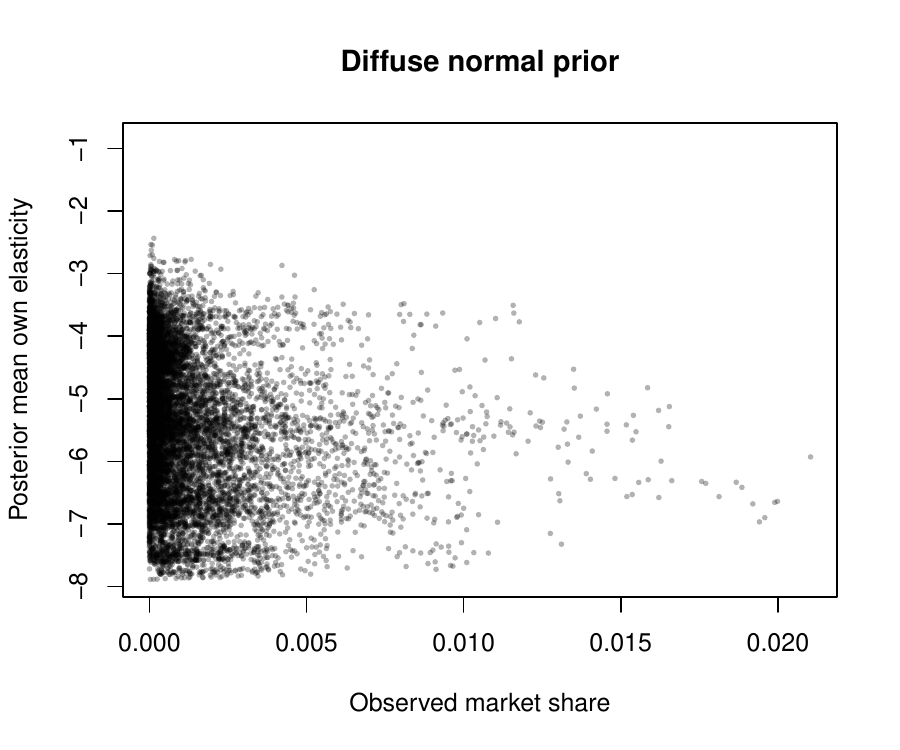}
\caption*{(b) Diffuse prior}
\end{minipage}

\caption{
Posterior mean own-price elasticities plotted against observed market shares.
}
\label{fig:at_elasticity_share}
\end{figure}

\FloatBarrier
\paragraph{Own-price elasticities and small-share products.}
Figure~\ref{fig:at_elasticity_share} plots posterior mean own-price elasticities against observed market shares. Under the diffuse prior, elasticities are larger in magnitude and more dispersed; as in the yogurt application, the left tail is more extreme than under shrinkage.

\section{Additional Simulation Results}\label{sec:appendix_simulation}
\FloatBarrier
\begin{table}[ht]
\caption{Simulation results of DGP3 and DGP4}
\label{tab:simulation_results_dense}
\centering
\resizebox{0.95\textwidth}{!}{%
\begin{threeparttable}

\begin{subtable}{\textwidth}
\begin{tabular}{c c c|c c c c|c c c c|c c c c|c c c c}
 & &  & \multicolumn{4}{c|}{BLP (infeasible, with cost IV)} & \multicolumn{4}{c|}{BLP (without cost IV)} & \multicolumn{4}{c|}{Shrinkage} & \multicolumn{4}{c}{Diffuse} \\
$J$ & $T$ &  & $\beta_p$ & $\beta_w$ & $\sigma$ & $\xi$ & $\beta_p$ & $\beta_w$ & $\sigma$ & $\xi$ & $\beta_p$ & $\beta_w$ & $\sigma$ & $\xi$ & $\beta_p$ & $\beta_w$ & $\sigma$ & $\xi$ \\
\hline
\multirow{2}{*}{5} & \multirow{2}{*}{10} & Bias & 0.034 & -0.640 & 0.245 & 0.966 & 0.026 & -0.613 & -0.153 & 0.981 & 0.119 & -0.134 & 0.066 & 0.208 & -0.653 & -0.559 & 0.555 & 0.951 \\
 & & SD & 0.515 & 0.045 & 0.822 & 1.000 & 0.865 & 0.217 & 1.473 & 1.130 & 0.416 & 0.253 & 0.606 & 0.439 & 0.605 & 0.280 & 0.852 & 1.188 \\
\hline
\multirow{2}{*}{5} & \multirow{2}{*}{50} & Bias & 0.083 & -0.636 & 0.127 & 0.962 & 0.245 & -0.549 & -1.030 & 0.976 & 0.049 & -0.079 & 0.197 & 0.116 & -3.627 & -1.257 & 10.079 & 0.922 \\
 & & SD & 0.530 & 0.019 & 0.759 & 0.989 & 1.775 & 0.506 & 1.221 & 1.448 & 0.571 & 0.140 & 0.709 & 0.257 & 3.281 & 0.382 & 2.932 & 3.171 \\
\hline
\multirow{2}{*}{15} & \multirow{2}{*}{10} & Bias & -0.067 & -0.637 & 0.170 & 0.960 & 0.568 & -0.751 & -0.461 & 0.981 & -0.088 & 0.018 & 0.116 & 0.010 & -0.836 & -0.765 & 1.755 & 0.912 \\
 & & SD & 0.570 & 0.044 & 0.498 & 1.001 & 1.454 & 0.410 & 1.524 & 1.391 & 0.527 & 0.116 & 0.477 & 0.279 & 1.154 & 0.449 & 1.202 & 1.368 \\
\hline
\multirow{2}{*}{15} & \multirow{2}{*}{50} & Bias & -0.041 & -0.636 & 0.072 & 0.960 & 0.930 & -0.746 & -1.242 & 0.979 & -0.054 & -0.025 & 0.123 & 0.053 & -2.933 & -0.927 & 6.384 & 0.786 \\
 & & SD & 0.511 & 0.025 & 0.392 & 1.000 & 2.677 & 0.798 & 0.960 & 1.958 & 0.515 & 0.052 & 0.408 & 0.241 & 2.167 & 0.282 & 2.374 & 2.212 \\
\hline
\end{tabular}
\caption*{(a) DGP3/dense exogenous case}
\end{subtable}

\vspace{0.5em}

\begin{subtable}{\textwidth}
\begin{tabular}{c c c|c c c c|c c c c|c c c c|c c c c}
 & &  & \multicolumn{4}{c|}{BLP (infeasible, with cost IV)} & \multicolumn{4}{c|}{BLP (without cost IV)} & \multicolumn{4}{c|}{Shrinkage} & \multicolumn{4}{c}{Diffuse} \\
$J$ & $T$ &  & $\beta_p$ & $\beta_w$ & $\sigma$ & $\xi$ & $\beta_p$ & $\beta_w$ & $\sigma$ & $\xi$ & $\beta_p$ & $\beta_w$ & $\sigma$ & $\xi$ & $\beta_p$ & $\beta_w$ & $\sigma$ & $\xi$ \\
\hline
\multirow{2}{*}{5} & \multirow{2}{*}{10} & Bias & 0.087 & -0.619 & -0.137 & 0.966 & 0.169 & -0.586 & -0.649 & 0.981 & 0.289 & -0.183 & -0.039 & 0.213 & -0.713 & -0.683 & 1.021 & 0.973 \\
 & & SD & 0.594 & 0.046 & 0.590 & 0.995 & 0.693 & 0.158 & 1.062 & 1.100 & 0.509 & 0.273 & 0.468 & 0.467 & 0.684 & 0.319 & 0.877 & 1.287 \\
\hline
\multirow{2}{*}{5} & \multirow{2}{*}{50} & Bias & 0.075 & -0.627 & -0.014 & 0.962 & 0.061 & -0.543 & -0.968 & 0.980 & 0.189 & -0.120 & 0.180 & 0.097 & -3.453 & -1.354 & 10.769 & 0.933 \\
 & & SD & 0.497 & 0.023 & 0.506 & 0.991 & 1.769 & 0.484 & 1.267 & 1.468 & 0.538 & 0.129 & 0.506 & 0.306 & 3.108 & 0.391 & 3.430 & 3.225 \\
\hline
\multirow{2}{*}{15} & \multirow{2}{*}{10} & Bias & -0.038 & -0.633 & 0.159 & 0.960 & 0.314 & -0.602 & -0.757 & 0.979 & 0.107 & -0.063 & 0.333 & 0.023 & -0.634 & -0.860 & 2.074 & 0.940 \\
 & & SD & 0.549 & 0.035 & 0.560 & 1.001 & 1.330 & 0.344 & 1.289 & 1.310 & 0.523 & 0.153 & 0.607 & 0.347 & 1.025 & 0.444 & 1.245 & 1.426 \\
\hline
\multirow{2}{*}{15} & \multirow{2}{*}{50} & Bias & 0.011 & -0.632 & 0.105 & 0.962 & 0.291 & -0.500 & -1.492 & 0.981 & 0.153 & -0.074 & 0.195 & 0.051 & -2.179 & -1.111 & 6.673 & 0.791 \\
 & & SD & 0.545 & 0.021 & 0.451 & 1.002 & 2.022 & 0.590 & 0.055 & 1.693 & 0.544 & 0.076 & 0.474 & 0.287 & 2.289 & 0.313 & 3.588 & 2.207 \\
\hline
\end{tabular}
\caption*{(b) DGP4/dense endogenous case}
\end{subtable}

\begin{tablenotes}
\item \footnotesize{\textit{Notes:} For $\beta_p$, $\beta_w$, and $\sigma$, the Bias rows report Monte Carlo average bias and the SD rows report Monte Carlo standard deviations of the point estimates. For $\xi$, the Bias rows report the average bias of estimated demand shocks and the SD rows report the average RMSE of estimated demand shocks across product--market pairs. In DGP3 and DGP4, product--market deviations are dense and normalized to have zero within-market mean and standard deviation $\sigma_\eta=1/3$. DGP3 uses exogenous prices. DGP4 introduces endogeneity by setting $\alpha_{jt}=\rho_\alpha \eta_{jt}/\sigma_\eta$ with $\rho_\alpha=0.3$. BLP (with cost IV) uses $(1,w_{jt},w_{jt}^2,u_{jt},u_{jt}^2)$ as instruments. BLP (without cost IV) uses $(1,w_{jt},w_{jt}^2,w_{jt}^3,w_{jt}^4)$ as instruments. Shrinkage corresponds to \texttt{rclogit\_par.stan}; Diffuse corresponds to \texttt{rclogit\_par\_normal.stan}. 
50 repeated experiments. 
}
\end{tablenotes}
\end{threeparttable}%
}
\end{table}
In DGP3 and DGP4, we consider dense product--market demand deviations. For each market \(t\), we first draw latent shocks 
$\tilde{\eta}_{jt}\sim N(0,1), \ j=1,\ldots,J$.
We then normalize these shocks within each market by defining 
$\eta_{jt}^{\ast}
=
\sigma_{\eta}
\frac{\tilde{\eta}_{jt}-\bar{\tilde{\eta}}_t}{s_{\tilde{\eta},t}}$,
where 
$\bar{\tilde{\eta}}_t
=
\frac{1}{J}\sum_{j=1}^{J}\tilde{\eta}_{jt}$ 
and \(s_{\tilde{\eta},t}\) is the within-market standard deviation of \(\tilde{\eta}_{jt}\). This normalization ensures that the product--market demand deviations sum to zero within each market and have comparable dispersion across designs with different values of \(J\). We set \(\sigma_{\eta}=1/3\). Unlike DGP1 and DGP2, in which only a subset of products have nonzero deviations, DGP3 and DGP4 generate nonzero demand deviations for all products and therefore provide dense misspecification designs for evaluating the shrinkage estimator.

DGP3 is the dense exogenous-price design. The unobserved demand component is given by 
$\xi_{jt}^{\ast}
=
\bar{\xi}_{t}^{\ast}
+
\eta_{jt}^{\ast}$,
with 
$\bar{\xi}_{t}^{\ast}=-1$, 
where \(\eta_{jt}^{\ast}\) is generated from the dense normalized design above. Price is generated as 
$p_{jt}
=
0.3w_{jt}+u_{jt}$, 
where \(u_{jt}\sim N(0,0.7^2)\) is an exogenous cost shock. Thus, conditional on \(w_{jt}\), price is independent of the latent demand deviation \(\eta_{jt}^{\ast}\).

DGP4 is the dense endogenous-price design. The demand shocks \(\eta_{jt}^{\ast}\) are generated in the same way as in DGP3, but price is allowed to depend on the latent demand deviation. Specifically, we set
$\alpha_{jt}^{\ast}
=
\rho_{\alpha}
\frac{\eta_{jt}^{\ast}}{\sigma_{\eta}}$,
with \(\rho_{\alpha}=0.3\), and generate price according to 
$p_{jt}
=
\alpha_{jt}^{\ast}
+
0.3w_{jt}
+
u_{jt}$. 
Because \(\alpha_{jt}^{\ast}\) is increasing in \(\eta_{jt}^{\ast}\), products with favorable latent demand shocks tend to have higher prices. DGP4 therefore creates a positive correlation between price \(p_{jt}\) and the unobserved demand component \(\xi_{jt}^{\ast}\), while maintaining a dense structure for product--market demand deviations. 

The dense designs provide a misspecification check for the sparsity assumption. In DGP3 and DGP4, all product--market deviations are nonzero, so the true latent demand component is not sparse. Nevertheless, the shrinkage estimator remains stable. Across both dense designs, shrinkage delivers small bias for the price coefficient and substantially lower RMSE for the latent demand shocks than the BLP and diffuse-prior estimators. This suggests that the shrinkage prior acts as a useful regularizer even when exact sparsity is violated.

The contrast with the diffuse prior is especially informative. Although the diffuse prior allows dense product--market deviations, it produces large distortions in the price coefficient and in the random-coefficient standard deviation, particularly in the larger-\(T\) designs. This indicates that unrestricted latent demand flexibility can absorb share variation in a way that destabilizes the structural parameters. 

BLP with the infeasible cost instruments estimates the price coefficient reasonably well, but performs poorly in recovering the latent demand shocks. BLP without cost instruments is less stable, consistent with the weak-instrument motivation of the simulation. Overall, the dense designs show that shrinkage does not rely mechanically on exact sparsity: it continues to discipline the high-dimensional latent demand component and yields stable structural estimates under a moderately dense structure.

\end{document}